\documentclass[traditabstract,onecolumn]{aa}
\usepackage{amsmath}
\usepackage{natbib}
\bibpunct{(}{)}{;}{a}{}{,}
\usepackage{textcomp}
\usepackage[T1]{fontenc}
\usepackage{txfonts}
\providecommand*\cramped[1]{{\kern-\nulldelimiterspace\radical0{#1}}}
\DeclareMathOperator{\Rot}{Rot}
\DeclareMathOperator{\Diag}{Diag}
\newcommand{\card}{\mathop{\#}\mathopen{}}
\newcommand{\indic}{\mathbf{1}}
\newcommand{\sep}{{,}\,}
\def\IE[#1,#2]{\llbracket#1{,}\,\mathopen{}#2\rrbracket} 
\newcommand{\BASE}{B}
\newcommand{\base}{u}
\newcommand{\Assoc}{A}
\newcommand{\COORD}{C}
\newcommand{\COORDpi}{\COORD'_{\smash[t]{i}}}
\newcommand{\rhopi}{\rho'_{\smash[t]{i}}}
\newcommand{\npi}{n'_{\smash[t]{i}}}
\newcommand{\coord}{c}
\newcommand{\sto}{{\text{s:o}}}
\newcommand{\ots}{{\text{o:s}}}
\newcommand{\loc}{{\text{loc}}}
\newcommand{\oto}{{\text{o:o}}}
\newcommand{\Lh}{L}
\newcommand{\Lhsto}{\Lh_\sto}
\newcommand{\Lhoto}{\Lh_\oto}
\newcommand{\Prob}{P}
\newcommand{\Ploc}{\Prob_{\!\loc}}
\newcommand{\Psto}{\Prob_{\!\sto}}
\newcommand{\Poto}{\Prob_{\!\oto}}
\newcommand{\fc}{f}
\newcommand{\sigmatot}{\mathring\sigma}
\newcommand{\nutot}{\mathring\nu}
\newcommand{\urpj}{\vec{u_{r'_{\smash[t]{j}}}}}
\newcommand{\udpj}{\vec{u_{\delta'_{\smash[t]{j}}}}}
\newcommand{\uapj}{\vec{u_{\alpha'_{\smash[t]{j}}}}}
\newcommand{\deltapj}{\delta'_{\smash[t]{j}}}
\newcommand{\alphapj}{\alpha'_{\smash[t]{j}}}
\newcommand{\coordpj}{\coord'_{\smash[t]{j}}}
\newcommand{\vrpj}{\vec{r'_{\smash[t]{j}}}}
\newcommand{\vrzi}{\vec{r^0_{\smash[t]{i}}}}
\newcommand{\vrpzj}{\vec{r'^0_{\smash[t]{j}}}}
\newcommand{\betapj}{\beta'_{\smash[t]{j}}}
\newcommand{\gammapj}{\gamma'_{\smash[t]{j}}}
\newcommand{\thetapj}{\theta'_{\smash[t]{j}}}
\newcommand{\Gammapj}{\Gamma'_{\smash[t]{j}}}
\newcommand{\gauss}{G}
\newcommand{\Left}{\mathopen{}\mathclose\bgroup\left}
\newcommand{\Right}{\aftergroup\egroup\right}
\newcommand{\leftsubstack}[1]{\subarray{l}#1\endsubarray}
\newcommand{\Stot}{S}
\newcommand{\nI}{n}
\newcommand{\NJi}{N'_{\smash[t]{i}}}
\newcommand{\nJ}{\cramped{n'}}
\newcommand{\KI}{K}
\newcommand{\KJ}{K'}
\newcommand{\MI}{M}
\newcommand{\MJ}{M'}
\newcommand{\MJj}{\MJ_{\smash[t]{j}}}
\newcommand{\MJk}{\MJ_{\smash[t]{k}}}
\newcommand{\MJl}{\MJ_{\smash[t]{\ell}}}
\newcommand{\Si}{S\!_i}
\renewcommand{\*}{\mathclose{}\,\mathopen{}}
\newcommand{\Dvc}{\ensuremath{D_{25}}}
\newcommand{\dvc}{\ensuremath{d_{25}}}
\newcommand{\meilleurj}{\check\jmath_i}
\newcommand{\df}{\mathrm{d}}
\newcommand{\LR}{\ensuremath{\textsc{\large lr}}}
\newcommand{\IRAS}{\textsc{\large iras}}
\newcommand{\LEDA}{\textsc{\large leda}}
\newcommand{\SED}{\textsc{\large sed}}
\newcommand{\transpose}[1]{#1^{\textsf{\textsc{t}}}}%
\newcommand{\chgbase}{T}
\begin{document}
\title{Probabilistic positional association\\
of astrophysical sources between catalogs}
\author{Michel Fioc}
\institute{Institut d'astrophysique de Paris, UPMC~- univ.~Paris~6, CNRS, 
  UMR 7095, 98bis boulevard Arago, F-75014 Paris, France
}
\offprints{M.~Fioc, \email{Michel.Fioc@iap.fr}}
\date{Received / Accepted}
\abstract{
  We describe a simple probabilistic method to cross-identify astrophysical
  sources from different catalogs and provide the probability that a source
  is associated with a source from another catalog or that it
  has no counterpart. When the positional
  uncertainty in one of the catalog is unknown, this method may be used
  to derive its typical value and even to study its dependence
  on the size of objects. It may also be applied when the true centers
  of a source and of its counterpart at another wavelength do not coincide.

  We extend this method to the case when there are only one-to-one associations
  between the catalogs.
}
\titlerunning{Probabilistic positional association of astrophysical sources}
\authorrunning{M. Fioc}
\maketitle
\section{Introduction}
The problem of cross-identifying sources between two catalogs 
$\KI$ and $\KJ$ 
has previously been studied by 
\citet{Condon}, \citet{DeRuiter}, \citet{Prestage}, \citet{SS} and 
\citet{Rutledge}, among others.
As evidenced by recent papers of \citet{BS}
and \citet{Pineau}, this field is still very active and will be more so with 
the wealth of forthcoming multiwavelength data.
Usually, the association is 
performed using a ``likelihood ratio'': this quantity is typically computed as 
the ratio of the probability of finding, 
at some distance from a source $\MI_i \in \KI$, a source $\MJj \in \KJ$,
if $\MJj$ is a counterpart of $\MI_i$,
to the probability that $\MJj$ is a chance association at the same position, 
given the local surface density of 
$\KJ$-sources. As noticed by \citet{SS}, there has been some confusion 
in the 
definition and interpretation of the likelihood ratio,
and, more importantly, in the estimation of the probability%
\footnote{%
  E.g., \citet{DeRuiter} state that, if there is 
  a counterpart, the closest object is always the right one, which is obviously
  wrong.%
}
that 
a source in $\KJ$ is the counterpart of a source in 
$\KI$. 

When associating sources from catalogs at different wavelengths,
some authors include in this likelihood ratio some \emph{a priori}
information on the spectral energy distribution (\SED) of the source.
As this work began, our primary goal was to build template observational \SED's
of galaxies from the optical to the far-infrared for different types of 
galaxies. 
We initially intended to cross-identify the \IRAS\ Faint Source Survey
\citep{FSS, FSC} with the \LEDA\ database \citep{LEDA}.
Because of the large positional inaccuracy of \IRAS\ data, 
special care was needed to identify optical sources with infrared ones.
While \IRAS\ data are by now quite outdated and have been superseded by  
Spitzer observations, 
we still think that the procedure we developed at that time
may be valuable for other studies.
Because we aimed to fit synthetic \SED's to the template observational
ones, we could not and did not want to make assumptions on the \SED\ of sources
based on their type, since this would have biased the procedure.
We therefore rely in what follows only on the positions to associate sources 
between catalogs.

The method we use is essentially similar to that of
\citet{SS}. Because thinking in terms of probabilities rather than of 
likelihood ratios highlights some implicit assumptions, we found it however 
useful for the sake of clarity to detail
hereafter our calculations; this allows us moreover to extend our work
to a case not covered by papers cited above (see Sect.~\ref{one}). 

We define our notations and explicit our general assumptions in 
Sect.~\ref{notations}.
In Sect.~\ref{several}, we compute the probability of association
under the assumption that a $\KI$-source has at most one counterpart
in $\KJ$ but that several $\KI$-sources may have the same counterpart
(``several-to-one'' associations).
We moreover determine the fraction of sources with a counterpart
and, if unknown, estimate the uncertainty on the position in one of the 
catalogs.
In Sect.~\ref{one}, we compute the probability of association under the 
assumption
that a $\KI$-source has at most one counterpart in $\KJ$ and that no other
$\KI$-source has the same counterpart (``one-to-one'' associations).
We provide in Sect.~\ref{practical} some guidance to help the user to 
implement these results.
The probability distribution of the relative positions
of associated sources is modeled in App.~\ref{cov}.
\section{Notations and general assumptions}
\label{notations}
We consider two catalogs $\KI$ and $\KJ$ defined on a common area $\Stot$
of the sky and use the following notations:
\begin{itemize}
\item
  $\card E$: number of elements of any set $E$;
\item
  $\MI_1$, \textellipsis, $\MI_\nI$, with $\nI \equiv \card\KI$:
  sources in $\KI$; 
\item
  $\MJ_{\smash[t]{1}}$, \textellipsis, $\MJ_{\smash[t]{\nJ}}$, with $\nJ \equiv \card\KJ$:
  sources in $\KJ$.  
\end{itemize}

We define the following events:
\begin{itemize}
\item
  $\coord_i$: $\MI_i$ is in the infinitesimal surface
  element $\df^2\vec{r_i}$ located at $\vec{r_i}$;
\item
  $\coordpj$: $\MJj$ is in the surface
  element $\df^2\vrpj$ located at $\vrpj$;
\item
  $\COORD \equiv \bigcap_{i=1}^\nI \coord_i$: the coordinates of all
  $\KI$-sources are known;
\item
  $\COORD' \equiv \bigcap_{j=1}^{\nJ} \coordpj$: the coordinates of all
  $\KJ$-sources are known;
\item 
  $\Assoc_{i\sep j}$, with $j>0$: $\MJj$ is the counterpart of $\MI_i$;
\item 
  $\Assoc_{i\sep 0}$: $\MI_i$ has no counterpart in $\KJ$, i.e.
  $\Assoc_{i\sep 0} = \overline{\bigcup_{j>0}\Assoc_{i\sep j}}$, where $\overline{\omega}$
  denotes the negation of any event $\omega$;
\item
  $\Assoc_{0\sep j}$: $\MJj$ has no counterpart in $\KI$.
\end{itemize}

We also write $\fc$
the \emph{a priori} probability $\Prob(\bigcup_{j>0} \Assoc_{i\sep j})$ that an element
of $\KI$ has a counterpart in $\KJ$ 
(so, $\Prob(\Assoc_{i\sep 0}) = 1-\fc$); we will see in Sects.~\ref{fractionsto}
and~\ref{fractionoto}
how to estimate $\fc$.
We moreover assume that any $\MI_i$ has at most
one counterpart in $\KJ$: $\Assoc_{i\sep j} \cap \Assoc_{i\sep k} = \varnothing$ 
if $j\neq k$.

Clustering is neglected in all the paper.
\section{Several-to-one associations}
\label{several}
In this section,
we do not make any assumption on the number of $\KI$-sources
that may be the counterpart of a given source of $\KJ$: this is a 
reasonable hypothesis if the angular resolution in $\KJ$ (e.g.\ \IRAS) 
is much poorer than in $\KI$ (e.g.\ \LEDA), since, in that case, 
several distinct objects of $\KI$ may be 
confused in $\KJ$. As evidenced by Sect.~\ref{local}, this is also the
assumption implicitly made by most of the authors cited in the introduction.
We call this the ``several-to-one'' case.
\subsection{Probability of association: all-sky computation}
\label{all-sky}
We want to compute%
\footnote{For the sake of clarity, let us mention that 
  we adopt the same decreasing order of precedence 
  of operators as in \emph{Mathematica} \citep{Mathematica}: 
  $\times$ and $/$; $\prod$; $\sum$; $+$ and $-$.}%
, in the several-to-one case,
the probability $\Psto(\Assoc_{i\sep j} \mid \COORD \cap \COORD')$ 
of association of sources $\MI_i$ and $\MJj$ ($j > 0$) 
or the probability that $\MI_i$ has no counterpart
($j = 0$), knowing the coordinates of all the objects in $\KI$ and $\KJ$. 
Remembering that, for any events $\omega_1$, $\omega_2$ and $\omega_3$, 
$\Prob(\omega_1 \mid \omega_2) = 
\Prob(\omega_1 \cap \omega_2)/\Prob(\omega_2)$ and
$\Prob(\omega_1 \cap \omega_2 \mid \omega_3) = 
\Prob(\omega_1 \mid \omega_2 \cap \omega_3) \* \Prob(\omega_2 \mid \omega_3)$, 
we have
\begin{align}
  \label{Pstodef}
  \Psto(\Assoc_{i\sep j} \mid \COORD \cap \COORD') 
  = \frac{\Psto(\Assoc_{i\sep j} \cap \COORD \cap \COORD')}{\Psto(\COORD \cap \COORD')}
  = \frac{\Psto(\Assoc_{i\sep j} \cap \COORD \mid \COORD')}{\Psto(\COORD \mid \COORD')}.
\end{align}

We first compute
$\Psto(\COORD \mid \COORD')$. Using the symbol $\biguplus$ for mutually 
exclusive events instead of $\bigcup$, we obtain
\begin{align}
  \Psto(\COORD \mid \COORD')
  &= \Psto\Bigl(\COORD \cap 
  \biguplus_{j_1=0}^{\nJ}\biguplus_{j_2=0}^{\nJ}\cdots\biguplus_{j_{\nI}=0}^{\nJ}
  \bigcap_{k=1}^\nI \Assoc_{k\sep j_k}
  \Bigm| \COORD'
  \Bigr) 
  = \sum_{j_1=0}^{\nJ} \sum_{j_2=0}^{\nJ}
  \cdots \sum_{j_{\nI}=0}^{\nJ} 
  \Psto\Bigl(\COORD 
  \cap \bigcap_{k=1}^{\nI} \Assoc_{k\sep j_k}
  \Bigm| \COORD'
  \Bigr) \notag \\
  &= \sum_{j_1=0}^{\nJ} \sum_{j_2=0}^{\nJ}
  \cdots \sum_{j_{\nI}=0}^{\nJ} 
  \Psto\Bigl(\COORD 
  \Bigm| \bigcap_{k=1}^{\nI} \Assoc_{k\sep j_k} 
  \cap \COORD'
  \Bigr) \* 
  \Psto\Bigl(\bigcap_{k=1}^{\nI} \Assoc_{k\sep j_k}
  \bigm| \COORD'
  \Bigr).
  \label{PstoCCpgen}
\end{align}

One has
\begin{align}
  \Psto\Bigl(\COORD \Bigm| \bigcap_{k=1}^{\nI} \Assoc_{k\sep j_k} \cap \COORD'\Bigr)
  &= \Psto\Bigl(\coord_1 \Bigm| \bigcap_{k=2}^{\nI} \coord_k 
  \cap \bigcap_{k=1}^{\nI} \Assoc_{k\sep j_k} \cap \COORD'\Bigr)
  \* \Psto\Bigl(\bigcap_{k=2}^{\nI} \coord_k 
  \Bigm| \bigcap_{k=1}^{\nI} \Assoc_{k\sep j_k} \cap 
  \COORD' \Bigr) 
  \notag \\
  &= \prod_{\ell=1}^{\nI}
  \Psto\Bigl(\coord_\ell \Bigm| \bigcap_{k=\ell+1}^{\nI} \coord_k 
  \cap \bigcap_{k=1}^{\nI} \Assoc_{k\sep j_k} \cap \COORD'\Bigr) 
  \label{prod_c_cp_A}
\end{align}
by iteration.

If $j_\ell \neq 0$, since $\MI_\ell$ is associated with $\MJ_{\smash[t]{j_\ell}}$ only,
\begin{align}
  \Psto\Bigl(\coord_\ell \Bigm| \bigcap_{k=\ell+1}^{\nI} \coord_k 
  \cap \bigcap_{k=1}^{\nI} \Assoc_{k\sep j_k} \cap \COORD'\Bigr)
  = \Psto(\coord_\ell \mid \Assoc_{\ell\sep j_\ell} \cap \coord'_{\smash[t]{j_\ell}})
  = \xi_{\ell\sep j_\ell} \* \df^2\vec{r_\ell},
  \label{jl_non_nul}
\end{align}
where
\[
\xi_{\ell\sep j_\ell} \equiv 
\frac{\exp\Bigl(-\frac{1}{2}\*\vec{\transpose r_{\smash[t]{\ell\sep j_\ell}}} \cdot
  \Gamma_{\smash[t]{\ell\sep j_\ell}}^{-1} \cdot \vec{r_{\ell\sep j_\ell}}\Bigr)}
{2\*\pi\*(\det \Gamma_{\ell\sep j_\ell})^{1/2}},
\]
$\vec{r_{\ell\sep j_\ell}} \equiv \vec{r'_{\smash[t]{j_\ell}}} - \vec{r_\ell}$
and the covariance matrix $\Gamma_{\ell\sep j_\ell}$ of $\vec{r_{\ell\sep j_\ell}}$ 
is computed as detailed in App.~\ref{cov}.
(Note that, in the several-to-one case considered here,
the computation of $\Psto(\COORD \mid \COORD')$ 
is easier than that of
$\Psto(\COORD' \mid \COORD)$:
because several $\MI_\ell$ may be associated with the same $\MJk$, the latter
would require to calculate
$\Psto(\coord'_{\smash[t]{k}} \mid \bigcap_{\ell=1{;}\, j_\ell=k}^\nI 
{[\coord_\ell \cap \Assoc_{\ell\sep j_\ell}]})$.
This does not matter in the one-to-one case studied in Sect.~\ref{one}.)

If $j_\ell = 0$, since $\MI_\ell$ is not associated with any source in $\KJ$
and clustering is neglected,
\begin{align}
  \Psto\Bigl(\coord_\ell \Bigm| \bigcap_{k=\ell+1}^{\nI} \coord_k \cap 
  \bigcap_{k=1}^{\nJ} \coord'_{\smash[t]{k}}
  \cap \bigcap_{k=1}^{\nI} \Assoc_{k\sep j_k}\Bigr)
  = \Psto(\coord_\ell \mid \Assoc_{\ell\sep 0})
  = \xi_{\ell\sep 0}\*\df^2\vec{r_\ell},
  \label{jl_nul}
\end{align}
where $\xi_{\ell\sep 0} \equiv 1/\Stot$ if we assume a uniform distribution 
of $\KI$-sources without counterpart as prior.

From Eqs.~\eqref{prod_c_cp_A}, \eqref{jl_non_nul} and~\eqref{jl_nul}, 
it follows that
\begin{align}
  \label{prod_xi_sto}
  \Psto\Bigl(\COORD \Bigm| \bigcap_{k=1}^{\nI} \Assoc_{k\sep j_k} \cap \COORD' \Bigr) &= 
  \lambda \* \prod_{k=1}^{\nI} \xi_{k\sep j_k},
\end{align}
where $\lambda \equiv \prod_{k=1}^{\nI} \df^2\vec{r_k}$.

We now compute $\Psto(\bigcap_{k=1}^{\nI} \Assoc_{k\sep j_k} \mid \COORD')$.
Without any other assumption,
$\Psto(\bigcap_{k=1}^{\nI} \Assoc_{k\sep j_k} \mid \COORD') = 
\Psto(\bigcap_{k=1}^{\nI} \Assoc_{k\sep j_k}) $. 
Let 
$m \equiv \card \{j_k > 0{;}\,\allowbreak k \in \IE[1, \nI]\}$. 
Since a given $\MJl$
may be the counterpart of several $\MI_k$ (i.e.\ the events 
$(\Assoc_{k\sep j_k})_{k\in\IE[1,\nI]}$ are independent whatever the values of the indices
$j_k$),
\[
\Psto\Bigl(\bigcap_{k=1}^{\nI} \Assoc_{k\sep j_k}\Bigr) = \prod_{k=1}^{\nI} \Psto(\Assoc_{k\sep j_k}).
\]
As $\Psto(\Assoc_{k\sep 0}) = 1-\fc$ and $\Psto(\Assoc_{k\sep j_k}) = \fc/\nJ$ for $j_k > 0$,
\begin{align}
  \label{PstoA}
  \Psto\Bigl(\bigcap_{k=1}^{\nI} \Assoc_{k\sep j_k}\Bigr) 
  = \Biggl(\frac{\fc}{\nJ}\Biggr)^m\*(1-\fc)^{\nI-m}.
\end{align}

Hence, from Eqs.~\eqref{Pstodef}, \eqref{PstoCCpgen}, 
\eqref{prod_xi_sto} and~\eqref{PstoA},
\begin{align}
  \label{PstoCCp}
  \Psto(\COORD \mid \COORD') 
  = \lambda\*\sum_{j_1=0}^{\nJ} \sum_{j_2=0}^{\nJ} \cdots \sum_{j_{\nI}=0}^{\nJ} 
  {\Biggl(\frac{\fc}{\nJ}\Biggr)^m\*(1-\fc)^{\nI-m}
  \*\prod_{k=1}^{\nI} \xi_{k\sep j_k}}
  = \lambda\*\Lhsto,
\end{align}
where
\begin{align}
  \label{Lh_sto}
  \Lhsto
  \equiv \sum_{j_1=0}^{\nJ}\sum_{j_2=0}^{\nJ} \cdots 
  \sum_{j_{\nI}=0}^{\nJ}\prod_{k=1}^\nI\zeta_{k\sep j_k} 
  = \prod_{k=1}^{\nI}\sum_{j_k=0}^{\nJ}\zeta_{k\sep j_k}
\end{align}
is the likelihood to observe 
the $\KI$-sources at their positions if the positions of $\KJ$-sources 
are known, $\zeta_{k\sep 0} \equiv (1-\fc)\*\xi_{k\sep 0}$
and $\zeta_{k\sep j_k} \equiv \fc\*\xi_{k\sep j_k}/\nJ$ if $j_k > 0$.

The computation of $\Psto(\Assoc_{i\sep j} \cap \COORD \mid \COORD')$ is similar to that of 
$\Psto(\COORD \mid \COORD')$:
\begin{align}
  \Psto(\Assoc_{i\sep j} \cap \COORD \mid \COORD') 
  &= \Psto\Bigl(\COORD 
  \cap \Assoc_{i\sep j} \cap 
  \biguplus_{j_1=0}^{\nJ} \cdots \biguplus_{j_{i-1}=0}^{\nJ}
  \biguplus_{j_{i+1}=0}^{\nJ} \cdots \biguplus_{j_{\nI}=0}^{\nJ}
  \bigcap_{\substack{k=1\\ k\neq i}}^{\nI} \Assoc_{k\sep j_k}
  \Bigm| \COORD'
  \Bigr) 
  = \Psto\Bigl(\COORD
  \cap \biguplus_{j_1=0}^{\nJ} \cdots \biguplus_{j_{i-1}=0}^{\nJ}
  \biguplus_{j_{i+1}=0}^{\nJ} \cdots \biguplus_{j_{\nI}=0}^{\nJ}
  \bigcap_{k=1}^{\nI} \Assoc_{k\sep j_k}
  \Bigm| \COORD'
  \Bigr) 
  \notag \\
  &= \sum_{j_1=0}^{\nJ}
  \cdots\sum_{j_{i-1}=0}^{\nJ}
  \sum_{j_{i+1}=0}^{\nJ}
  \cdots \sum_{j_{\nI}=0}^{\nJ} 
  \Psto\Bigl(\COORD 
  \Bigm| \bigcap_{k=1}^{\nI} \Assoc_{k\sep j_k}
  \cap \COORD'
  \Bigr) \* 
  \Psto\Bigl(\bigcap_{k=1}^{\nI} \Assoc_{k\sep j_k}
  \Bigm| \COORD'
  \Bigr),
  \label{PStoACCpgen}
\end{align}
where we have put $j_i \equiv j$.

Let $m^\ast \equiv \card \{j_k > 0{;}\, k \in \IE[1, \nI]\}$
(indices $j_k$ are those of Eq.~\eqref{PStoACCpgen}).
As for 
$\Psto(\COORD \mid \COORD')$, 
\begin{align}
  \Psto(\Assoc_{i\sep j} \cap \COORD \mid \COORD') &= 
  \lambda\*\sum_{j_1=0}^{\nJ} \cdots
  \sum_{j_{i-1}=0}^{\nJ} \sum_{j_{i+1}=0}^{\nJ}
  \cdots \sum_{j_{\nI}=0}^{\nJ}  
  {\Biggl(\frac{\fc}{\nJ}\Biggr)^{m^\ast}\*(1-\fc)^{\nI-m^\ast}
    \prod_{k=1}^{\nI} \xi_{k\sep j_k}} 
  = \lambda\*\zeta_{i\sep j_i}\*\sum_{j_1=0}^{\nJ}\cdots
  \sum_{j_{i-1}=0}^{\nJ}\sum_{j_{i+1}=0}^{\nJ}\cdots
  \sum_{j_{\nI}=0}^{\nJ}\prod_{\substack{k=1 \\ k\neq i}}^\nI \zeta_{k\sep j_k}
  \notag \\
  &= \lambda\*\zeta_{i\sep j}\*\prod_{\substack{k=1 \\ k\neq i}}^{\nI}
  \sum_{j_k=0}^{\nJ}\zeta_{k\sep j_k}.
  \label{PstoACCp}
\end{align}

Finally, from Eqs.~\eqref{Pstodef}, \eqref{PstoCCp}, \eqref{Lh_sto} 
and~\eqref{PstoACCp},
\begin{align}
  \label{Psto}
  \Psto(\Assoc_{i\sep j} \mid \COORD \cap \COORD') &= \frac{
    \zeta_{i\sep j}\*\prod_{\leftsubstack{k=1 \\ k\neq i}}^{\nI}
    \sum_{j_k=0}^{\nJ}\zeta_{k\sep j_k}
  }{
    \prod_{k=1}^{\nI}\sum_{j_k=0}^{\nJ}\zeta_{k\sep j_k}
  } 
  = \frac{\zeta_{i\sep j}}{\sum_{k=0}^{\nJ}\zeta_{i\sep k}}
  \\
  &= \label{Psto2}
  \Left\{
  \begin{aligned}
    \frac{\fc\*\xi_{i\sep j}}{
      (1-\fc)\*\nJ\!/\Stot + \fc\*\sum_{k=1}^{\nJ}\xi_{i\sep k}} 
    & \quad \text{if } j > 0,\\
    \frac{(1-\fc)\*\nJ\!/\Stot}{
      (1-\fc)\*\nJ\!/\Stot + \fc\*\sum_{k=1}^{\nJ}\xi_{i\sep k}}
    & \quad \text{if } j = 0.
    \end{aligned}
    \Right.
\end{align}

The probability $\Psto(\Assoc_{0\sep j} \mid \COORD \cap \COORD')$ that $\MJj$ 
has no counterpart in $\KI$ can be computed in this way:
\begin{align*}
  \Psto(\Assoc_{0\sep j} \cap \COORD \mid \COORD')
  &= \Psto\Bigl(\COORD \cap \Assoc_{0\sep j} \cap 
  \biguplus_{j_1=0}^{\nJ} \biguplus_{j_2=0}^{\nJ}
  \cdots \biguplus_{j_{\nI}=0}^{\nJ}
  \bigcap_{k=1}^{\nI} \Assoc_{k\sep j_k}
  \Bigm| \COORD'\Bigr) 
  = \Psto\Bigl(\COORD \cap 
  \biguplus_{\substack{j_1=0 \\ j_1\neq j}}^{\nJ} 
  \biguplus_{\substack{j_2=0\\ j_2\neq j}}^{\nJ}
  \cdots \biguplus_{\substack{j_{\nI}=0\\ j_\nI\neq j}}^{\nJ}
  \bigcap_{k=1}^{\nI} \Assoc_{k\sep j_k}
  \Bigm| \COORD'\Bigr) \\
  &= \sum_{\substack{j_1=0 \\ j_1\neq j}}^{\nJ} 
  \sum_{\substack{j_2=0\\ j_2\neq j}}^{\nJ}
  \cdots \sum_{\substack{j_{\nI}=0\\ j_\nI\neq j}}^{\nJ}
  \Psto\Bigl(\COORD \cap   
  \bigcap_{k=1}^{\nI} \Assoc_{k\sep j_k}
  \Bigm| \COORD'\Bigr) 
  = \lambda\*\sum_{\substack{j_1=0 \\ j_1\neq j}}^{\nJ} 
  \sum_{\substack{j_2=0\\ j_2\neq j}}^{\nJ}
  \cdots \sum_{\substack{j_{\nI}=0\\ j_\nI\neq j}}^{\nJ} 
  \prod_{k=1}^{\nI}\zeta_{k\sep j_k} 
  = \lambda\*\prod_{k=1}^{\nI}\sum_{\substack{j_k=0\\ j_k\neq j}}^{\nJ}\zeta_{k\sep j_k}
\end{align*}
and
\begin{align}
  \Psto(\Assoc_{0\sep j} \mid \COORD \cap \COORD') &= 
  \frac{\Psto(\Assoc_{0\sep j} \cap \COORD \mid \COORD')}{\Psto(\COORD \mid \COORD')}
  = \frac{\lambda\*\prod_{k=1}^{\nI}\sum_{\leftsubstack{j_k=0\\ j_k\neq j}}^{\nJ}
    \zeta_{k\sep j_k}}{\lambda\*\prod_{k=1}^{\nI}\sum_{j_k=0}^{\nJ}\zeta_{k\sep j_k}}
  = \prod_{k=1}^{\nI}\frac{\sum_{j_k=0}^{\nJ}\zeta_{k\sep j_k} - \zeta_{k\sep j}}{
    \sum_{j_k=0}^{\nJ}\zeta_{k\sep j_k}}
  = \prod_{k=1}^{\nI}{\Biggl(1-\frac{\zeta_{k\sep j_k}}{\sum_{j_k=0}^{\nJ}\zeta_{k\sep j_k}}\Biggr)}
  \notag \\
  & = \prod_{k=1}^{\nI}{(1 - \Psto[\Assoc_{k\sep j} \mid \COORD \cap \COORD'])}.
  \label{PstoA0j}
\end{align}
\subsection{Fraction of sources with a counterpart and other unknown 
  parameters}
\label{fractionsto}
\subsubsection{Estimates}
Besides $\fc$, the probabilities $\Prob(\Assoc_{i\sep j} \mid \COORD \cap \COORD')$ may 
depend on other unknown parameters, 
e.g.\ $\sigmatot$ and $\nutot$ (cf.~App.~\ref{cov}). Let us write them
$x_1$, $x_2$, etc., and $\vec x \equiv (x_1, x_2, \ldots)$.
An estimate $\hat{\vec x}$ of $\vec x$ may be obtained by maximizing the
likelihood $\Lh$
with respect to $\vec x$ (and with the constraint $\hat\fc_{\sto} \in[0, 1]$), or, 
equivalently, by finding the solution 
$\hat{\vec x}$ of
\begin{align}
  \label{max_Lh}
  \frac{\partial\ln\Lh}{\partial\vec x} = 0.
\end{align}

For any parameter $x_p$, as all the $\zeta_{i\sep j}$ are strictly positive
and $\ln\Lhsto = \sum_{i=1}^{\nI} \ln\sum_{k=0}^{\nJ}\zeta_{i\sep k}$ 
(Eq.~\eqref{Lh_sto}),
\begin{align}
  \frac{\partial\ln\Lhsto}{\partial x_p}
  &= \sum_{i=1}^{\nI}\frac{\partial\ln\sum_{k=0}^{\nJ}\zeta_{i\sep k}}{\partial x_p}
  = \sum_{i=1}^{\nI}\sum_{j=0}^{\nJ}\frac{\partial\zeta_{i\sep j}/\partial x_p}{
    \sum_{k=0}^{\nJ}\zeta_{i\sep k}}
  = \sum_{i=1}^{\nI}\sum_{j=0}^{\nJ}\frac{\partial\ln\zeta_{i\sep j}}{\partial x_p}\*
  \frac{\zeta_{i\sep j}}{\sum_{k=0}^{\nJ}\zeta_{i\sep k}} \notag \\
  &= \sum_{i=1}^{\nI}\sum_{j=0}^{\nJ}\frac{\partial\ln\zeta_{i\sep j}}{\partial x_p}\*
  \Psto(\Assoc_{i\sep j} \mid \COORD \cap \COORD').
  \label{der_Lh_sto}
\end{align}

Let us consider in particular the case $x_p = \fc$. Note that
$\partial\ln\zeta_{i\sep 0}/\partial\fc = -1/(1-\fc)$
and $\partial\ln\zeta_{i\sep j}/\partial\fc = 1/\fc$ for $j > 0$.
Since $\sum_{j=0}^{\nJ} \Psto(\Assoc_{i\sep j} \mid \COORD \cap \COORD') = 1$,
\begin{align}
  \sum_{j=0}^{\nJ}\frac{\partial\ln\zeta_{i\sep j}}{\partial x_p}\*
  \Psto(\Assoc_{i\sep j} \mid \COORD \cap \COORD')
  &= -\frac{\Psto(\Assoc_{i\sep 0} \mid \COORD \cap \COORD')}{1-\fc}
  + \sum_{j=1}^{\nJ} \frac{\Psto(\Assoc_{i\sep j} \mid \COORD \cap \COORD')}{\fc}
  = -\frac{\Psto(\Assoc_{i\sep 0} \mid \COORD \cap \COORD')}{1-\fc}
  + \frac{1-\Psto(\Assoc_{i\sep 0} \mid \COORD \cap \COORD')}{\fc} \notag \\
  &= \frac{(1-\fc) - \Psto(\Assoc_{i\sep 0} \mid \COORD \cap \COORD') }{
    \fc\*(1-\fc)}.
  \label{somme_j}
\end{align}
Summing on $i$, we obtain
\begin{align}
  \frac{\partial\ln\Lhsto}{\partial\fc}
  &= \frac{\nI\*(1-\fc) - \sum_{i=1}^{\nI}\Psto(\Assoc_{i\sep 0} \mid \COORD \cap \COORD')}{
    \fc\*(1-\fc)}.
\end{align}
So, as expected, an estimate of the probability that a source in $\KI$ has a
counterpart in $\KJ$ is given by
\begin{align}
  \hat\fc_{\sto} &= 1 - \frac{1}{\nI}\*
  \sum_{i=1}^{\nI}\expandafter\hat\Psto(\Assoc_{i\sep 0} \mid \COORD \cap \COORD'),
  \label{f_est} \\
  &= \frac{1}{\nI}\*
  \sum_{i=1}^{\nI}\sum_{j=1}^{\nJ}\expandafter\hat\Psto(\Assoc_{i\sep j} \mid \COORD \cap \COORD').
  \label{f_est2}
\end{align}
Note that, since $\partial^2\zeta_{i\sep j}/\partial\fc^2 = 0$ for all 
$(i, j) \in \IE[1, \nI]\times\IE[0, \nJ]$,
\begin{align}
  \label{concave}
  \frac{\partial^2\ln\Lhsto}{\partial\fc^2} =
  -\sum_{i=1}^\nI{\Biggl(\frac{
      \sum_{j=0}^{\nJ}\partial\zeta_{i\sep j}/\partial\fc
    }{
      \sum_{j=0}^{\nJ}\zeta_{i\sep j}}\Biggr)^2} < 0
\end{align}
for all $\fc$, 
so $\partial\ln\Lhsto/\partial\fc$ has at most one zero in $[0, 1]$:
$\hat\fc_{\sto}$ is unique.

One may also compute an estimate of the fraction $\fc'$ of $\KJ$-sources 
with a counterpart from
\begin{align}
  \hat\fc'_{\sto} = 1 - \frac{1}{\nJ}\*
  \sum_{j=1}^{\nJ}\expandafter\hat\Psto(\Assoc_{0\sep j} \mid \COORD \cap \COORD').
  \label{f'}
\end{align}
One can easily check from Eqs.~\eqref{f_est2}, \eqref{f'} and~\eqref{PstoA0j}
that $\hat\fc_{\sto}/\nJ > \hat\fc'_{\sto}/\nI$ in the several-to-one case.
\subsubsection{Uncertainties}
It may be interesting to know the uncertainties on the unknown parameters.
For large numbers of sources, the covariance matrix 
$V$ of $\hat{\vec x}$ is asymptotically given by 
\begin{align}
  \label{cov_x}
  \Bigl(V^{-1}\Bigr)_{p\sep q} 
  = \Biggl(-\frac{\partial^2\ln\Lh}{\partial x_p\*\partial x_q}
  \Biggr)_{\vec x=\hat{\vec x}}
\end{align}
\citep{KS}.

Let us write with a circumflex accent all the quantities calculated at
$\vec x = \hat{\vec x}$.
From
\[
\frac{\partial^2\ln\Lh}{\partial x_p\*\partial x_q}
= \frac{1}{\expandafter \Prob(\COORD \mid \COORD')}\*\frac{\partial^2 \Prob(\COORD \mid \COORD')}{
  \partial x_p\*\partial x_q} - \frac{1}{\Prob^2(\COORD \mid \COORD')}
\*\frac{\partial \Prob(\COORD \mid \COORD')}{\partial x_p}\*
\frac{\partial \Prob(\COORD \mid \COORD')}{\partial x_q},
\]
one obtains
\begin{align}
  \label{der2_Lh}
  \frac{\hat\partial^2\ln\Lh}{\hat\partial x_p\*\hat\partial x_q}
  = \frac{1}{\expandafter\expandafter\hat \Prob(\COORD \mid \COORD')}\*
  \frac{\hat\partial^2 \Prob(\COORD \mid \COORD')}{
    \hat\partial x_p\*\hat\partial x_q}.
\end{align}
One has
\begin{align}
  \label{der2_Lh2}
  \frac{\partial^2 \Psto(\COORD \mid \COORD')}{\partial x_p\*\partial x_q} 
  = \sum_{i=1}^{\nI}\sum_{j=0}^{\nJ}{\frac{\partial^2\ln\zeta_{i\sep j}}{
    \partial x_p\*\partial x_q}\*\Psto(\Assoc_{i\sep j} \cap \COORD \mid \COORD')}
  + \sum_{i=1}^{\nI}\sum_{j=0}^{\nJ}{\frac{\partial\ln\zeta_{i\sep j}}{
    \partial x_p}\*\frac{\partial \Psto(\Assoc_{i\sep j} \cap \COORD \mid \COORD')}{
    \partial x_q}}.
\end{align}
For any product of strictly positive functions $g_k$ of some variable $y$,
\begin{align}
  \label{der_prod}
  \frac{\partial\prod_{k=1}^{\nI} g_k}{\partial y} 
  = \sum_{i=1}^{\nI}{\frac{\partial g_i}{\partial y}\*
    \prod_{\substack{k=1\\ k\neq i}}^{\nI} g_k}
  = \sum_{i=1}^{\nI}{\frac{\partial\ln g_i}{\partial y}\*
    \prod_{k=1}^{\nI} g_k},
\end{align}
so, using Eq.~\eqref{PstoACCp},
\begin{align}
  \frac{\partial \Psto(\Assoc_{i\sep j} \cap \COORD \mid \COORD')}{\partial x_q}
  &= \lambda\*\frac{\partial\zeta_{i\sep j}}{\partial x_q}\*
  \prod_{\substack{k=1\\ k\neq i}}^{\nI}\sum_{j_k=0}^{\nJ}\zeta_{k\sep j_k}
  + \lambda\*\zeta_{i\sep j}\*\sum_{\substack{\ell=1\\\ell\neq i}}^{\nI}{
    \frac{\partial\sum_{j_\ell=0}^{\nJ}\zeta_{\ell\sep j_\ell}}{\partial x_q}\*
    \prod_{\substack{k=1\\ k\not\in\{i{,}\,\ell\}
      }}^{\nI}\sum_{j_k=0}^{\nJ}\zeta_{k\sep j_k}} \notag \\
  &= \lambda\*\frac{\partial\ln\zeta_{i\sep j}}{\partial x_q}\*\zeta_{i\sep j}\*
  \prod_{\substack{k=1\\ k\neq i}}^{\nI}\sum_{j_k=0}^{\nJ}\zeta_{k\sep j_k}
  + \lambda\*\frac{\zeta_{i\sep j}}{\sum_{j_i=0}^{\nJ}\zeta_{i\sep j_i}}\*
  \sum_{\substack{\ell=1\\\ell\neq i}}^{\nI}\sum_{j_\ell=0}^{\nJ}
  {\frac{\partial\ln\zeta_{\ell\sep j_\ell}}{\partial x_q}\*
    \zeta_{\ell\sep j_\ell}\*
    \prod_{\substack{k=1\\ k\neq\ell}}^{\nI}\sum_{j_k=0}^{\nJ}\zeta_{k\sep j_k}} \notag \\
  &= \frac{\partial\ln\zeta_{i\sep j}}{\partial x_q}\*
  \Psto(\Assoc_{i\sep j} \cap \COORD \mid \COORD')
  + \Psto(\Assoc_{i\sep j} \mid \COORD \cap \COORD')\*
  \sum_{\substack{\ell=1\\\ell\neq i}}^{\nI}\sum_{j_\ell=0}^{\nJ}
  {\frac{\partial\ln\zeta_{\ell\sep j_\ell}}{\partial x_q}\*
    \Psto(\Assoc_{\ell\sep j_\ell} \cap \COORD \mid \COORD')}.
  \label{der_P}
\end{align}
For $\vec x = \hat{\vec x}$, 
\begin{align}
  \sum_{\substack{\ell=1\\\ell\neq i}}^{\nI}\sum_{j_\ell=0}^{\nJ}
  {\frac{\hat\partial\ln\zeta_{\ell\sep j_\ell}}{\hat\partial x_q}\*
    \expandafter\hat \Psto(\Assoc_{\ell\sep j_\ell} \cap \COORD \mid \COORD')}
  &= \sum_{\ell=1}^{\nI}\sum_{j_\ell=0}^{\nJ}
  {\frac{\hat\partial\ln\zeta_{\ell\sep j_\ell}}{\hat\partial x_q}\*
    \expandafter\hat \Psto(\Assoc_{\ell\sep j_\ell} \cap \COORD \mid \COORD')}
  - \sum_{j_i=0}^{\nJ}
  {\frac{\hat\partial\ln\zeta_{i\sep j_i}}{\hat\partial x_q}\*
    \expandafter\hat \Psto(\Assoc_{i\sep j_i} \cap \COORD \mid \COORD')} \notag \\
  &= - \sum_{j_i=0}^{\nJ}
  {\frac{\hat\partial\ln\zeta_{i\sep j_i}}{\hat\partial x_q}\*
    \expandafter\hat \Psto(\Assoc_{i\sep j_i} \cap \COORD \mid \COORD')}
  \label{der_P2}
\end{align}
since the first term on the right-hand side of the first line
is zero from Eq.~\eqref{der_Lh_sto}.
Finally, combining Eqs.~\eqref{der2_Lh2}, \eqref{der_P}, \eqref{der_P2} 
and dividing by 
$\expandafter\hat \Psto(\COORD \mid \COORD')$, 
we obtain
\begin{align}
  \label{der2_Lh_sto}
  \frac{\hat\partial^2\ln\Lhsto}{\hat\partial x_p\*\hat\partial x_q}
  &= \sum_{i=1}^{\nI}\sum_{j=0}^{\nJ}{\Biggl(\frac{\hat\partial^2\ln\zeta_{i\sep j}}{
    \hat\partial x_p\*\hat\partial x_q} + 
  \frac{\hat\partial\ln\zeta_{i\sep j}}{\hat\partial x_p}\*
  \frac{\hat\partial\ln\zeta_{i\sep j}}{\hat\partial x_q}\Biggr)
  \*\expandafter\hat \Psto(\Assoc_{i\sep j} \mid \COORD \cap \COORD')} \notag \\
  &\mathrel{\phantom{=}}{}- \sum_{i=1}^{\nI}{\Biggl(
  \sum_{j=0}^{\nJ}{\frac{\hat\partial\ln\zeta_{i\sep j}}{\hat\partial x_p}
  \*\expandafter\hat \Psto[\Assoc_{i\sep j} \mid \COORD \cap \COORD']}\Biggr) \* 
  \sum_{j=0}^{\nJ}{\frac{\hat\partial\ln\zeta_{i\sep j}}{\hat\partial x_q}
  \*\expandafter\hat \Psto(\Assoc_{i\sep j} \mid \COORD \cap \COORD')}}.
\end{align}

In particular, for $x_p = x_q = \fc$,
$\partial^2\ln\zeta_{i\sep j}/\partial\fc^2 + 
(\partial\ln\zeta_{i\sep j}/\partial\fc)^2= 0$, whether $j=0$ or not.
From Eqs.~\eqref{somme_j} and~\eqref{f_est},
\begin{align}
  \label{der2_Lh_sto2}
  \frac{\hat\partial^2\ln\Lhsto}{\hat\partial\fc^2}
  &= -\sum_{i=1}^{\nI}{\Biggl(\frac{1}{\hat\fc_{\sto}} - 
  \frac{\expandafter\hat\Psto[\Assoc_{i\sep 0} \mid \COORD \cap \COORD']}{
    \hat\fc_{\sto}\*(1-\hat\fc_{\sto})}\Biggr)^2}
  \notag \\
  &= \frac{\nI}{\hat\fc_{\sto}^2} - 
  \frac{\sum_{i=1}^{\nI} \expandafter\hat \Psto^2(\Assoc_{i\sep 0} \mid \COORD \cap \COORD')}{
    \hat\fc_{\sto}^2\*(1-\hat\fc_{\sto})^2}.
\end{align}

\subsection{Probability of association: local computation}
\label{local}
In the several-to-one case, a purely local computation of the probability
of association between a given $\MI_i$ and some $\MJj$ ($j > 0$), or of the
probability that $\MI_i$ has no counterpart in $\KJ$, is also possible. 

Let us consider a region $D_i$
of area $\Si$ containing the position of $\MI_i$, and 
such that we can safely hypothesize that the $\KJ$-counterpart of $\MI_i$, 
if any,
will be inside. 
We assume that the local surface density $\rhopi$ 
of $\KJ$-sources unrelated to $\MI_i$ is uniform on 
$D_i$. 
To avoid biasing the estimate if $\MI_i$ has a counterpart,
$\rhopi$ may be computed from the number of $\KJ$-sources in 
a region surrounding, but not overlapping, $D_i$.

Besides the $\Assoc_{i\sep j}$, we consider the following events: 
\begin{itemize}
\item 
  $\NJi$: $D_i$ contains $\npi$ sources;
\item 
  $\COORDpi \equiv \bigcap_{j \in I_i} \coordpj$,
  where $I_i \equiv \{j \mid \MJj \in D_i\}$.
\end{itemize}

We want to compute the probability that a source $\MJj$ in $D_i$ is the 
counterpart of $\MI_i$, given the positions of the 
neighbors, i.e.
$\Ploc(\Assoc_{i\sep j} \mid \COORDpi\cap \NJi)$. 
We have
\begin{align*}
  \Ploc(\Assoc_{i\sep j} \mid \COORDpi\cap \NJi)
  &= \frac{\Ploc(\Assoc_{i\sep j} \cap \COORDpi\cap \NJi)}{
    \Ploc(\COORDpi \cap \NJi)}
  = \frac{\Ploc(\COORDpi \cap \Assoc_{i\sep j} \cap \NJi)}{
    \Ploc(\COORDpi \cap \biguplus_{k\in I_i\cup\{0\}} \Assoc_{i\sep k} \cap \NJi)} \\
  &= \frac{\Ploc(\COORDpi \cap \Assoc_{i\sep j} \cap \NJi)}{
    \sum_{k\in I_i\cup\{0\}} \Ploc(\COORDpi \cap \Assoc_{i\sep k} \cap \NJi)}
  = \frac{\Ploc(\COORDpi \mid \Assoc_{i\sep j} \cap \NJi)\*\Ploc(\Assoc_{i\sep j} 
    \cap \NJi)}{\sum_{k\in I_i\cup\{0\}} \Ploc(\COORDpi\mid \Assoc_{i\sep k} \cap \NJi)
    \*\Ploc(\Assoc_{i\sep k} \cap \NJi)} \\
  &=\frac{\Ploc(\COORDpi\mid \Assoc_{i\sep j} \cap \NJi)\*\Ploc(\Assoc_{i\sep j} \mid \NJi)}{
    \sum_{k\in I_i\cup\{0\}} \Ploc(\COORDpi\mid \Assoc_{i\sep k} \cap \NJi)
    \*\Ploc(\Assoc_{i\sep k} \mid \NJi)}.
\end{align*}

If $j>0$, $\Ploc(\Assoc_{i\sep j} \mid \NJi)=\Ploc(\bigcup_{k\in I_i}\Assoc_{i\sep k} \mid \NJi)/\npi$ (one sees here why
the event $\NJi$ was defined: otherwise, $\Ploc(\Assoc_{i\sep j})$ could not be computed as
$\Ploc(\bigcup_{k\in I_i}\Assoc_{i\sep k})/\npi$ because $\npi$ would be undefined). 
Now,
\[\Ploc\Bigl(\bigcup_{k\in I_i}\Assoc_{i\sep k} \mid \NJi\Bigr) = 
\frac{\Ploc( \NJi \cap \bigcup_{k\in I_i}\Assoc_{i\sep k})}{\Ploc(\NJi)} =
\frac{\Ploc(\NJi\mid \bigcup_{k\in I_i}\Assoc_{i\sep k})\*\Ploc(\bigcup_{k\in I_i}\Assoc_{i\sep k})}
{\Ploc(\NJi\mid \Assoc_{i\sep 0})\*\Ploc(\Assoc_{i\sep 0})+\Ploc(\NJi\mid \bigcup_{k\in I_i}\Assoc_{i\sep k})\*\Ploc(\bigcup_{k\in I_i}\Assoc_{i\sep k})}.\]
If clustering is negligible, the number of sources randomly distributed 
with a mean surface density
$\rhopi$ in an area $\Si$ follows a Poissonian distribution, so 
\[\Ploc\Bigl(\NJi\mid \bigcup_{k\in I_i}\Assoc_{i\sep k}\Bigr) = 
\frac{(\rhopi\*\Si)^{\npi-1}\*\exp(-\rhopi\*\Si)}{(\npi-1)!}
\quad\text{($\npi-1$ random sources in $\Si$)}\] 
and 
\[\Ploc(\NJi\mid \Assoc_{i\sep 0}) = 
\frac{(\rhopi\*\Si)^{\npi}\*\exp(-\rhopi\*\Si)}{\npi!}
\quad\text{($\npi$ random sources in $\Si$)}.\]
Thus, 
\[
\Ploc(\Assoc_{i\sep j} \mid \NJi) = 
\Left\{\begin{aligned}
  \frac{\fc}{\npi\*\fc+(1-\fc)\*\rhopi\*\Si} & \quad \text{if } j > 0,\\
  \frac{(1-\fc)\*\rhopi\*\Si}{\npi\*\fc+(1-\fc)\*\rhopi\*\Si} 
  & \quad \text{if } j = 0.
\end{aligned}\Right.
\]

For $j > 0$,
\[
\Ploc(\COORDpi\mid \Assoc_{i\sep j} \cap \NJi) =
\xi_{i\sep j}\*\df^2\vrpj\*
\prod_{\substack{k\in I_i\\ k\neq j}} \frac{\df^2\vec{r'_{\smash[t]{k}}}}{\Si}
\]
(rigorously, $\xi_{i\sep j}$ should be replaced by 
$\xi_{i\sep j}/\Ploc(\MJj\in D_i\mid \Assoc_{i\sep j})$, but 
$\Ploc(\MJj\not\in D_i \mid \Assoc_{i\sep j})$ is negligible), and
\[
\Ploc(\COORDpi\mid \Assoc_{i\sep 0}\cap \NJi) = 
\prod_{k\in I_i} \frac{\df^2\vec{r'_{\smash[t]{k}}}}{\Si}.
\]

Finally, 
\begin{equation}
  \label{cpsto}
  \Ploc(\Assoc_{i\sep j}\mid \COORDpi\cap \NJi)
  = \Left\{\begin{aligned}
      \frac{\fc\*\LR_{i\sep j}}{(1-\fc)+\fc\*\sum_{k\in I_i}\LR_{i\sep k}}
      & \quad \text{if } j > 0, \\
      \frac{(1-\fc)}{(1-\fc)+\fc\*\sum_{k\in I_i}\LR_{i\sep k}}
      & \quad \text{if } j = 0,
    \end{aligned}\Right.
\end{equation}
where $\LR_{i\sep k} \equiv \xi_{i\sep k}/\rhopi$ is the 
``likelihood ratio''.
\emph{Mutatis mutandis}, one obtains the same result as Eq.~(14) of 
\citet{Pineau} and aforementioned authors. When extended to the all sky
(i.e.\ $\Si \to \Stot$), $\rhopi$ is replaced by $\nJ\!/\Stot$ in 
Eq.~\eqref{cpsto}, $\sum_{k\in I_i}$ by $\sum_{k=1}^{\nJ}$ and
one recovers Eq.~\eqref{Psto2}.

The index $\meilleurj$ of the most likely counterpart 
$\MJ_{\smash[t]{\meilleurj}}$ of $\MI_i$ is the value 
of $j > 0$ maximizing $\LR_{i\sep j}$.
Usually, $\sum_{k=1;\,k\neq \meilleurj}^{\npi}\LR_{i\sep k}\ll \LR_{i\sep \meilleurj}$,
so
\[
\Psto(\Assoc_{i\sep \meilleurj}\mid \COORD \cap \COORD')
\approx \frac{\fc\*\LR_{i\sep \meilleurj}}{(1-\fc)+\fc\*\LR_{i\sep \meilleurj}}.
\]
As a ``poor man's'' recipe, if the value of $\fc$ is unknown and not too 
close to either $0$ or $1$,
an association may be considered as true if $\LR_{i\sep \meilleurj}\gg 1$ and as 
false if $\LR_{i\sep \meilleurj}\ll 1$.
Where to set the boundary between true associations and false ones is
somewhat arbitrary.
For a large sample, however, $\fc$ can be determined from the distribution 
of the positions of all the sources, as shown in Sect.~\ref{fractionsto}.
\section{One-to-one associations}
\label{one}
In Sect.~\ref{several}, a given $\MJj$ may be associated with several $\MI_i$:
the probabilities are actually asymmetric in $\MI_i$ and $\MJj$ 
and, while $\sum_{j=0}^{\nJ} \Psto(\Assoc_{i\sep j}\mid \COORD \cap \COORD') = 1$ for all $\MI_i$, 
one may well have 
$\sum_{i=1}^{\nI} \Psto(\Assoc_{i\sep j}\mid \COORD \cap \COORD') > 1$ 
for some sources $\MJj$.

Here, we assume not only that each $\KI$-source is associated with at most
one $\KJ$-source, but that each $\KJ$-source is associated with at most 
one $\KI$-source. We call this the ``one-to-one'' case and note $\Poto$
the probabilities calculated under this assumption. As far as we know and
despite some attempt by \citet{Rutledge},
this problem has not been solved previously.

Since a $\KJ$-potential counterpart of $\MI_i$ within some 
neighborhood $D_i$ of $\MI_i$
might in fact be the true counterpart
of another source $\MI_k$ outside of $D_i$, 
there is no obvious way to extend the exact local several-to-one 
computation of Sect.~\ref{local}
to the one-to-one case. We therefore have to consider either the whole sky,
as in Sect.~\ref{all-sky}, 
or at least some large enough region around both $\MI_i$ and $\MJj$ 
to neglect side effects.

In the case of one-to-one associations, a source of $\KI$ and a source of $\KJ$
play symmetrical roles; in particular, 
$\Poto(\Assoc_{i\sep j}) = \fc/\nJ = \fc'\!/\nI$.
However, for practical reasons (cf.~Eq.~\eqref{binom}), 
we name $\KI$ the catalog with
the fewer objects and $\KJ$ the other one, so $\nI \leqslant \nJ$ in the following.

\subsection{Probability of association}
We want to compute $\Poto(\Assoc_{i\sep j} \mid \COORD \cap \COORD')$ 
for $i > 0$. 
We still have
\begin{align}
  \label{Potodef}
  \Poto(\Assoc_{i\sep j} \mid \COORD \cap \COORD') 
  = \frac{\Poto(\Assoc_{i\sep j} \cap \COORD \mid \COORD')}{\Poto(\COORD \mid \COORD')}
\end{align}
and
\begin{align*}
  \Poto(\COORD \mid \COORD') 
  &=  \Poto\Bigl(\COORD \cap 
  \bigcup_{j_1=0}^{\nJ}\bigcup_{j_2=0}^{\nJ}\cdots\bigcup_{j_{\nI}=0}^{\nJ}
  \bigcap_{k=1}^\nI \Assoc_{k\sep j_k}
  \Bigm| \COORD'
  \Bigr).
\end{align*}
As $\Assoc_{i\sep j} \cap \Assoc_{k\sep \ell} = \varnothing$ if 
$i \neq k$ and $j = \ell > 0$, this reduces to
\begin{align*}
  \Poto(\COORD \mid \COORD') 
  &=  \Poto\Bigl(\COORD \cap
  \biguplus_{\substack{j_1=0 \\ j_1\not\in J_0}}^{\nJ}
  \biguplus_{\substack{j_2=0 \\ j_2\not\in J_1}}^{\nJ}
  \cdots 
  \biguplus_{\substack{j_{\nI}=0 \\ j_{\nI}\not\in J_{\nI-1}}}^{\nJ} 
  \bigcap_{k=1}^\nI \Assoc_{k\sep j_k}
  \Bigm| \COORD'
  \Bigr),
\end{align*}
where $J_0 \equiv \varnothing$ and $J_k$ is defined iteratively for all 
$k \in \IE[1, \nI]$ by $J_{k} \equiv (J_{k-1} \cup \{j_k\}) \setminus \{0\}$.
Hence,
\begin{align}
  \Poto(\COORD \mid \COORD') 
  &= \sum_{\substack{j_1=0 \\ j_1\not\in J_0}}^{\nJ} 
  \sum_{\substack{j_2=0 \\ j_2\not\in J_1}}^{\nJ}
  \cdots \sum_{\substack{j_{\nI}=0 \\ j_{\nI}\not\in J_{\nI-1}}}^{\nJ} 
  \Poto\Bigl(\COORD \cap 
  \bigcap_{k=1}^{\nI} \Assoc_{k\sep j_k}
  \Bigm| \COORD'
  \Bigr) \notag \\
  &= \sum_{\substack{j_1=0 \\ j_1\not\in J_0}}^{\nJ} 
  \sum_{\substack{j_2=0 \\ j_2\not\in J_1}}^{\nJ}
  \cdots \sum_{\substack{j_{\nI}=0 \\ j_{\nI}\not\in J_{\nI-1}}}^{\nJ} 
  \Poto\Bigl(\COORD 
  \Bigm| \bigcap_{k=1}^{\nI} \Assoc_{k\sep j_k}
  \cap \COORD'
  \Bigr) 
  \* \Poto\Bigl(\bigcap_{k=1}^{\nI} \Assoc_{k\sep j_k} 
  \Bigm| \COORD'
  \Bigr).  
  \label{PotoCCpgen}
\end{align}

As in the several-to-one case,
\begin{align}
  \label{prod_xi_oto}
  \Poto\Bigl(\COORD \Bigm| \bigcap_{k=1}^{\nI} \Assoc_{k\sep j_k} \cap \COORD'\Bigr) &= 
  \lambda\*\prod_{k=1}^{\nI} \xi_{k\sep j_k}.
\end{align}

We now have to compute 
$\Poto(\bigcap_{k=1}^{\nI} \Assoc_{k\sep j_k} \mid \COORD') = 
\Poto(\bigcap_{k=1}^{\nI} \Assoc_{k\sep j_k})$. 
Let $m \equiv \card J_{\nI}$ and 
$X$ be a random variable describing the number of associations between
$\KI$ and $\KJ$:
\[
\Poto\Bigl(\bigcap_{k=1}^{\nI} \Assoc_{k\sep j_k}\Bigr) = 
\Poto\Bigl(\bigcap_{k=1}^{\nI} \Assoc_{k\sep j_k} \Bigm| X = m\Bigr) \* \Poto(X = m) +
\Poto\Bigl(\bigcap_{k=1}^{\nI} \Assoc_{k\sep j_k} \Bigm| X \neq m\Bigr) \* \Poto(X \neq m).
\]
Since $\Poto(\bigcap_{k=1}^{\nI} \Assoc_{k\sep j_k} \mid X \neq m) = 0$, one just has
to compute $\Poto(\bigcap_{k=1}^{\nI} \Assoc_{k\sep j_k} \mid X = m)$ and
$\Poto(X = m)$.

There are $\nI!/(m!\*[\nI-m]!)$ choices of $m$ elements among $\nI$ in $\KI$,
and $\nJ!/(m!\*[\nJ-m]!)$ of $m$ elements among $\nJ$ in $\KJ$.
The number of permutations
of $m$ elements is $m!$, so
the total number of one-to-one associations of $m$ elements from $\KI$ to 
$m$ elements of $\KJ$ is 
\[
m!\*\frac{\nI!}{m!\*(\nI-m)!}\*\frac{\nJ!}{m!\*(\nJ-m)!}.
\]
The inverse of this number is
\begin{align}
  \label{PotoAm}
  \Poto\Bigl(\bigcap_{k=1}^{\nI} \Assoc_{k\sep j_k} \Bigm| X = m\Bigr) = 
  \frac{m!\*(\nI-m)!\*(\nJ-m)!}{\nI!\*\nJ!}.
\end{align}

With our definition of $\KI$ and $\KJ$, $\nI \leqslant \nJ$, so all
the elements of $\KI$ may have a counterpart in $\KJ$ jointly.
Therefore, $\Poto(X = m)$ is given by the binomial law:
\begin{align}
  \label{binom}
  \Poto(X = m) = \frac{\nI!}{m!\*(\nI-m)!}\*\fc^m\*(1-\fc)^{\nI-m}.
\end{align}

From Eqs.~\eqref{PotoCCpgen}, \eqref{prod_xi_oto}, \eqref{PotoAm} and \eqref{binom}, we obtain
\begin{align}
  \Poto(\COORD \mid \COORD') &= 
  \lambda\*\sum_{\substack{j_1=0 \\ j_1\not\in J_0}}^{\nJ} 
  \sum_{\substack{j_2=0 \\ j_2\not\in J_1}}^{\nJ}
  \cdots \sum_{\substack{j_{\nI}=0 \\ j_{\nI}\not\in J_{\nI-1}}}^{\nJ} 
  {\frac{(\nJ-m)!}{\nJ!}
    \* \fc^m \* (1-\fc)^{\nI-m}\*\prod_{k=1}^{\nI} \xi_{k\sep j_k}}
  \notag \\
  &= \lambda\*\Lhoto,
  \label{PotoCCp}
\end{align}
where
\begin{align}
  \label{Lh_oto}
  \Lhoto \equiv
  \sum_{\substack{j_1=0 \\ j_1\not\in J_0}}^{\nJ}
  \sum_{\substack{j_2=0 \\ j_2\not\in J_1}}^{\nJ}\cdots
  \sum_{\substack{j_{\nI}=0 \\ j_{\nI}\not\in J_{\nI-1}}}^{\nJ}
  \prod_{k=1}^\nI \eta_{k\sep j_k},
\end{align}
$\eta_{k\sep 0} \equiv \zeta_{k\sep 0}$ and
$\eta_{k\sep j_k} \equiv \fc\*\xi_{k\sep j_k}/(\nJ-\card J_{k-1})$ if $j_k > 0$.

$\Poto(\Assoc_{i\sep j} \cap \COORD \mid \COORD')$ is computed in the same way
as $\Poto(\COORD \mid \COORD')$:
\begin{align*}
  \Poto(\Assoc_{i\sep j} \cap \COORD \mid \COORD') 
  &= \Poto\Bigl(\COORD \cap 
  \Assoc_{i\sep j} \cap 
  \biguplus_{\substack{j_1=0 \\ j_1\not\in J^\ast_0}}^{\nJ} 
  \cdots 
  \biguplus_{\substack{j_{i-1}=0 \\ j_{i-1}\not\in J^\ast_{i-2}}}^{\nJ} 
  \biguplus_{\substack{j_{i+1}=0 \\ j_{i+1}\not\in J^\ast_{i}}}^{\nJ} 
  \cdots 
  \biguplus_{\substack{j_{\nI}=0 \\ j_{\nI}\not\in J^\ast_{\nI-1}}}^{\nJ} 
  \bigcap_{\substack{k=1\\ k\neq i}}^\nI \Assoc_{k\sep j_k}
  \Bigm| \COORD'
  \Bigr) \\
  &= \Poto\Bigl(\COORD \cap
  \biguplus_{\substack{j_1=0 \\ j_1\not\in J^\ast_0}}^{\nJ} 
  \cdots 
  \biguplus_{\substack{j_{i-1}=0 \\ j_{i-1}\not\in J^\ast_{i-2}}}^{\nJ} 
  \biguplus_{\substack{j_{i+1}=0 \\ j_{i+1}\not\in J^\ast_{i}}}^{\nJ} 
  \cdots 
  \biguplus_{\substack{j_{\nI}=0 \\ j_{\nI}\not\in J^\ast_{\nI-1}}}^{\nJ} 
  \bigcap_{k=1}^\nI \Assoc_{k\sep j_k}
  \Bigm| \COORD'
  \Bigr),
\end{align*}
where $j_i \equiv j$, $J^\ast_0 \equiv \{j\} \setminus \{0\}$ 
and $J^\ast_{k} \equiv (J^\ast_{k-1} \cup \{j_k\}) \setminus \{0\}$
for all $k \in \IE[1, \nI]$, so
\begin{align*}
  \Poto(\Assoc_{i\sep j} \cap \COORD \mid \COORD') 
  = \sum_{\substack{j_1=0 \\ j_1\not\in J^\ast_0}}^{\nJ}
  \cdots\sum_{\substack{j_{i-1}=0 \\ j_{i-1}\not\in J^\ast_{i-2}}}^{\nJ}
  \sum_{\substack{j_{i+1}=0 \\ j_{i+1}\not\in J^\ast_{i}}}^{\nJ}
  \cdots \sum_{\substack{j_{\nI}=0 \\ j_{\nI}\not\in J^\ast_{\nI-1}}}^{\nJ} 
  \Poto\Bigl(\COORD 
  \Bigm| \bigcap_{k=1}^{\nI} \Assoc_{k\sep j_k}
  \cap \COORD'
  \Bigr) \* 
  \Poto\Bigl(\bigcap_{k=1}^{\nI} \Assoc_{k\sep j_k}
  \Bigm| \COORD'
  \Bigr).
\end{align*}

Let $m^\ast \equiv \card J^\ast_{\nI}$. As for 
$\Poto(\COORD \mid \COORD')$,
\begin{align}
  \Poto(\Assoc_{i\sep j} \cap \COORD \mid \COORD') &= 
  \lambda\*\sum_{\substack{j_1=0 \\ j_1\not\in J^\ast_0}}^{\nJ} \cdots
  \sum_{\substack{j_{i-1}=0 \\ j_{i-1}\not\in J^\ast_{i-2}}}^{\nJ}
  \sum_{\substack{j_{i+1}=0 \\ j_{i+1}\not\in J^\ast_{i}}}^{\nJ}
  \cdots \sum_{\substack{j_{\nI}=0 \\ j_{\nI}\not\in J^\ast_{\nI-1}}}^{\nJ} 
  {\frac{(\nJ-m^\ast)!}{\nJ!}
    \* \fc^{m^\ast} \* (1-\fc)^{\nI-m^\ast}\*\prod_{k=1}^{\nI} \xi_{k\sep j_k}} 
  \notag \\
  &= \lambda\*\eta^\ast_{i\sep j}\*
  \sum_{\substack{j_1=0 \\ j_1\not\in J^\ast_0}}^{\nJ}
  \cdots
  \sum_{\substack{j_{i-1}=0 \\ j_{i-1}\not\in J^\ast_{i-2}}}^{\nJ}
  \sum_{\substack{j_{i+1}=0 \\ j_{i+1}\not\in J^\ast_{i}}}^{\nJ}
  \cdots 
  \sum_{\substack{j_{\nI}=0 \\ j_{\nI}\not\in J^\ast_{\nI-1}}}^{\nJ}
  \prod_{\substack{k=1\\ k\neq i}}^\nI \eta^\ast_{k\sep j_k},
  \label{PotoACCp}
\end{align}
where 
$
\eta^\ast_{k\sep j_k} \equiv 
\fc\*\xi_{k\sep j_k}/(\nJ-\card J^\ast_{k-1})
$
if $k\neq i$ and $j_k > 0$, and 
$\eta^\ast_{k\sep j_k} = \zeta_{k\sep j_k}$ otherwise.

Finally, from Eqs.~\eqref{Potodef}, \eqref{PotoCCp}, \eqref{Lh_oto} and~\eqref{PotoACCp},
\begin{align}
  \label{Poto}
  \Poto(\Assoc_{i\sep j} \mid \COORD \cap \COORD') 
  &= \frac{
    \zeta_{i\sep j}\*
    \sum_{\leftsubstack{j_1=0 \\ j_1\not\in J^\ast_0}}^{\nJ}
    \cdots
    \sum_{\leftsubstack{j_{i-1}=0 \\ j_{i-1}\not\in J^\ast_{i-2}}}^{\nJ}
    \sum_{\leftsubstack{j_{i+1}=0 \\ j_{i+1}\not\in J^\ast_{i}}}^{\nJ}
    \cdots 
    \sum_{\leftsubstack{j_{\nI}=0 \\ j_{\nI}\not\in J^\ast_{\nI-1}}}^{\nJ}
    \prod_{\leftsubstack{k=1\\ k\neq i}}^\nI \eta^\ast_{k\sep j_k}
  }{
    \sum_{\leftsubstack{j_1=0 \\ j_1\not\in J_0}}^{\nJ}
    \sum_{\leftsubstack{j_2=0 \\ j_2\not\in J_1}}^{\nJ}
    \cdots 
    \sum_{\leftsubstack{j_{\nI}=0 \\ j_{\nI}\not\in J_{\nI-1}}}^{\nJ}
    \prod_{k=1}^\nI \eta_{k\sep j_k}
  }.
\end{align}

The probability that a source $\MJj$ has no counterpart in $\KI$ is simply
given by
\[
\Poto(\Assoc_{0\sep j} \mid \COORD \cap \COORD') 
= 1-\sum_{k=1}^{\nI} \Poto(\Assoc_{k\sep j}\mid \COORD \cap \COORD').
\]
\subsection{Fraction of sources with a counterpart and other unknown 
  parameters}
\label{fractionoto}
\subsubsection{Estimates}
As in the several-to-one case, an estimate $\hat{\vec x}_{\oto}$ of the set $\vec x$ 
of unknown parameters may be obtained by solving Eq.~\eqref{max_Lh}
(with the constraint $\hat\fc_{\oto} \in [0, \nI/\nJ]$). 
As the number of terms in $\Lhoto$ grows exponentially with $\nI$ and $\nJ$,
Eq.~\eqref{Lh_oto} seems useless for this purpose.
Fortunately, the computation of 
$\Lhoto$ is not necessary if the probabilities
$\Poto(\Assoc_{i\sep j} \mid \COORD \cap \COORD')$ are known (we will see in
Sect.~\ref{impl_oto} how to approximate these).

Indeed, for any parameter $x_p$, let us show that we get the same result 
(Eq.~\eqref{der_Lh_sto}) as in the
several-to-one case.
Using Eq.~\eqref{der_prod}, we obtain
\begin{equation}
  \label{der_Lh_x_gauche}
  \frac{\partial \Poto(\COORD \mid \COORD')}{\partial x_p} 
  = \lambda \*
  \sum_{\substack{j_1=0\\j_1\notin J_0}}^{\nJ}\sum_{\substack{j_2=0\\j_2\notin J_1}}^{\nJ}
  \cdots\sum_{\substack{j_{\nI}=0\\j_{\nI}\notin J_{\nI-1}}}^{\nJ}\sum_{i=1}^{\nI}{
  \frac{\partial\ln\eta_{i\sep j_i}}{\partial x_p}\*
  \prod_{k=1}^{\nI}\eta_{k\sep j_k}}.
\end{equation}
The expression of 
$\Poto(\Assoc_{i\sep j} \cap \COORD \mid \COORD')$ 
may also be written
\[
\Poto(\Assoc_{i\sep j} \cap \COORD \mid \COORD')
= \lambda \*
\sum_{\substack{j_1=0\\j_1\notin J_0}}^{\nJ}\sum_{\substack{j_2=0\\j_2\notin J_1}}^{\nJ}
\cdots\sum_{\substack{j_{\nI}=0\\j_{\nI}\notin J_{\nI-1}}}^{\nJ}
{\indic(j_i = j)\*\prod_{k=1}^{\nI} \eta_{k\sep j_k}},
\]
where $\indic$ is the indicator function (i.e.\ $\indic(j_i = j) = 1$ if 
proposition ``$j_i = j$\kern1pt'' is true and $\indic(j_i = j) = 0$ otherwise),
so 
\begin{align}
  \label{der_Lh_x_droite}
  \sum_{i=1}^{\nI}\sum_{j=0}^{\nJ}{\frac{\partial\ln\zeta_{i\sep j}}{\partial x_p}
    \*\Poto(\Assoc_{i\sep j} \cap \COORD \mid \COORD')} 
  &= \lambda\*\sum_{i=1}^{\nI}\sum_{\substack{j_1=0\\j_1\notin J_0}}^{\nJ}
  \sum_{\substack{j_2=0\\j_2\notin J_1}}^{\nJ}
  \cdots\sum_{\substack{j_{\nI}=0\\j_{\nI}\notin J_{\nI-1}}}^{\nJ}\sum_{j=0}^{\nJ}
  {\indic(j_i = j)\*\frac{\partial\ln\zeta_{i\sep j}}{\partial x_p}\*
    \prod_{k=1}^{\nI}\eta_{k\sep j_k}} \notag \\
  &= \lambda\*\sum_{i=1}^{\nI}\sum_{\substack{j_1=0\\j_1\notin J_0}}^{\nJ}
  \sum_{\substack{j_2=0\\j_2\notin J_1}}^{\nJ}
  \cdots\sum_{\substack{j_{\nI}=0\\j_{\nI}\notin J_{\nI-1}}}^{\nJ}
  {\frac{\partial\ln\zeta_{i\sep j_i}}{\partial x_p}\*
    \prod_{k=1}^{\nI}\eta_{k\sep j_k}}.
\end{align}
If $j_i = 0$, $\eta_{i\sep j_i} = \zeta_{i\sep j_i}$; and if $j_i > 0$,
the numerators of $\eta_{i\sep j_i}$ and $\zeta_{i\sep j_i}$ are the same
and their denominators do not depend on $x_p$: 
in all cases, $\partial\ln\eta_{i\sep j_i}/\partial x_p 
= \partial\ln\zeta_{i\sep j_i}/\partial x_p$. 
The right-hand sides of Eqs.~\eqref{der_Lh_x_gauche} and~\eqref{der_Lh_x_droite}
are therefore identical. 
Dividing their left-hand sides by 
$\Poto(\COORD \mid \COORD')$, 
one obtains again
\begin{equation}
  \label{der_Lh_x}
  \frac{\partial\ln\Lhoto}{\partial x_p} 
  = \sum_{i=1}^{\nI}\sum_{j=0}^{\nJ}{\frac{\partial\ln\zeta_{i\sep j}}{
      \partial x_p}\*\Poto(\Assoc_{i\sep j} \mid \COORD \cap \COORD')}.
\end{equation}

For $x_p = \fc$, one still has
$\partial\ln\zeta_{i\sep 0}/\partial\fc = -1/(1-\fc)$ and
$\partial\ln\zeta_{i\sep j}/\partial\fc = 1/\fc$ if $j > 0$, so,
as in the several-to-one case,
\begin{align}
  \frac{\partial\ln\Lhoto}{\partial\fc} 
  &= \frac{\nI\*(1-\fc) 
    - \sum_{i=1}^{\nI} \Poto(\Assoc_{i\sep 0} \mid \COORD \cap \COORD')}{\fc\*(1-\fc)}.
\end{align}
\subsubsection{Uncertainties}
Regarding uncertainties on the $x_p$, Eqs.~\eqref{cov_x}, \eqref{der2_Lh} 
and \eqref{der2_Lh2} are valid in the one-to-one case too, 
so, from Eq.~\eqref{der_Lh_x},
\[
\frac{\hat\partial^2 \ln\Lhoto}{\hat\partial x_p\*\hat\partial x_q} 
= \sum_{i=1}^{\nI}\sum_{j=0}^{\nJ}{\frac{\hat\partial^2\ln\zeta_{i\sep j}}{
  \hat\partial x_p\*\hat\partial x_q}\*
\expandafter\hat\Poto(\Assoc_{i\sep j} \mid \COORD \cap \COORD')}
+ \sum_{i=1}^{\nI}\sum_{j=0}^{\nJ}{\frac{\hat\partial\ln\zeta_{i\sep j}}{
  \hat\partial x_p}\*\frac{\hat\partial \Poto(\Assoc_{i\sep j} \mid \COORD \cap \COORD')}{
  \hat\partial x_q}}.
\]
Contrary to the several-to-one case, no simple exact analytic 
expression of the terms 
$\hat\partial \Poto(\Assoc_{i\sep j} \mid \COORD \cap \COORD')/\hat\partial x_q$
could be obtained. These derivatives may be computed numerically 
using finite differences; however, unless the fraction of sources having several
likely counterparts is high, Eqs.~\eqref{der2_Lh_sto} and~\eqref{der2_Lh_sto2}
should provide a more convenient approximation of the covariance matrix of
$\hat{\vec x}_{\oto}$.
\section{Practical implementation}
\label{practical}
\subsection{Several-to-one case}
\subsubsection{Neighbors only!}
\label{neighbors}
In the several-to-one case, the computation of the probability of association 
$\Psto(\Assoc_{i\sep j} \mid \COORD \cap \COORD')$ between $\MI_i$ and $\MJj$
from Eq.~\eqref{Psto} 
is without problem if $\fc$ and the positional uncertainties
are known. However, the number of calculations for the whole sample
or for the determination of $\hat{\vec x}$ is of the order of 
$\nI\*\nJ{}^2
$. 

As $\zeta_{i\sep k}$ rapidly tends to $0$ when the angular distance $r_{i\sep k}$
between $\MI_i$ and $\MJk$ increases, there is no need to sum
from $k = 1$ to $\nJ$ in Eq.~\eqref{Psto}, nor to compute explicitly all the
$\Psto(\Assoc_{i\sep j} \mid \COORD \cap \COORD')$. If $R$ is some 
angular distance above which 
$\xi_{i\sep k} \ll \nJ\!/\Stot$, one may set $\xi_{i\sep k}$ to $0$ (and 
$\Psto(\Assoc_{i\sep k})$ too) if $r_{i\sep k} > R$ and replace the
sums $\sum_{k=1}^{\nJ}$ by 
$\sum_{k=1;\, r_{i\sep k}\leqslant R}^{\nJ}$. 

In fact, for most $\MI_i$,
one does not even need to test whether $r_{i\sep k} \leqslant R$ for each 
$\MJk \in \KJ$.
Let us write $E_i$ the domain of right ascensions $\alpha'$ out of which
no point $\MJ$ of declination $\delta'$ closer than $R$ to $\MI_i$ may be found.
The angular distance $\psi$ between $\MJ$ and $\MI_i$ is given (cf.~Eq.~\eqref{psi}) by
\[
\cos\psi
= \cos(\alpha'-\alpha_i)\*\cos\delta_i\*\cos\delta' + \sin\delta_i\*\sin\delta'.
\]

If $\delta_i \in [-\pi/2+R, \pi/2-R]$, the minimum of $\cos(\alpha'-\alpha_i)$ 
under the constraint $\cos\psi \geqslant \cos R$ is reached 
when $\sin\delta' = \sin\delta_i/{\cos R}$ and 
\[
\cos(\alpha'-\alpha_i) = \frac{\!\sqrt{\cos^2 R - \sin\mathclose{}^2\,\delta_i}}{\cos\delta_i}.
\]
Let $\Delta_i 
\equiv \arccos\Bigl(\!\sqrt{\cos^2 R - \sin\mathclose{}^2\,\delta_i}/{\cos\delta_i}\Bigr)$.
The domain $E_i$ is given by
\begin{align*}
  E_i = \Left\{
  \begin{aligned}
    &[0, \alpha_i + \Delta_i - 2\*\pi] \cup 
    [\alpha_i-\Delta_i, 2\*\pi] 
    && \text{if } \alpha_i + \Delta_i > 2\*\pi,  \\
    &[0, \alpha_i + \Delta_i] \cup 
    [\alpha_i-\Delta_i + 2\*\pi, 2\*\pi]
    && \text{if } \alpha_i - \Delta_i < 0, \\
    &[\alpha_i - \Delta_i, \alpha_i+\Delta_i]
    && \text{otherwise}.
  \end{aligned}
  \Right.
\end{align*}
If $\delta_i \in [-\pi/2, {-}\pi/2+R] \cup [\pi/2-R, \pi/2]$, one has
$E_i = [0, 2\*\pi]$.

For a catalog $\KJ$ ordered by increasing right ascension
(if not, this is the first thing to do), 
one may easily find the subset
of indices $k$ for which $\alpha'_{\smash[t]{k}} \in E_i$. For instance, 
if $E_i = [\alpha_i - \Delta_i, \alpha_i+\Delta_i]$,
one just has to find by dichotomy the indices $k^-$ and $k^+$ such that
$\alpha'_{\smash[t]{k^--1}} < \alpha_i - \Delta_i 
\leqslant \alpha'_{\smash[t]{k^-}}$
and $\alpha'_{\smash[t]{k^+}} \leqslant \alpha_i + \Delta_i 
< \alpha'_{\smash[t]{k^++1}}$. 
The sums $\sum_{k=1;\, r_{i\sep k}\leqslant R}^{\nJ}$ may then be replaced by 
$\sum_{k=k^-;\, r_{i\sep k}\leqslant R}^{k^+}$.

In all cases, the sum may be further restricted to sources with a declination
$\delta'_{\smash[t]{k}} \in [\delta_i-R, \delta_i+R] \cap [-\pi/2, \pi/2]$.
\subsubsection{Fraction of sources with a counterpart}
All the probabilities depend on $\fc$ and, possibly, other unknown parameters
like $\sigmatot$ and $\nutot$. These parameters may be found by solving 
Eq.~\eqref{max_Lh} using Eq.~\eqref{der_Lh_sto}. 

If the fraction of sources with a counterpart is the only unknown, 
the $\xi_{i\sep j}$ need to be computed only once and $\fc$
may 
be easily determined from Eq.~\eqref{f_est}.
Denote $g$ the function
\begin{align*}
  g\colon  [0, 1] & \to \mathbb{R}, \\
   \fc &\mapsto 1-\frac{1}{\nI}\*\sum_{i=1}^\nI\Psto(\Assoc_{i\sep 0} \mid \COORD \cap \COORD').
\end{align*}
Let us show that,
for any $\fc_0 \in \mathopen]0, 1\mathclose[$, the sequence
$(\fc_k)_{k\in\mathbb{N}}$ defined by $\fc_{k+1} \equiv g(\fc_k)$ tends to 
$\hat\fc$.

First, note that
\[
g(\fc) = \fc + \frac{\fc\*(1-\fc)}{\nI} \* \frac{\partial\ln\Lhsto}{
  \partial\fc}.
\]
The only fixed points of $g$ are hence $0$, $1$ and $\hat\fc$.
As $\partial^2\ln\Lhsto/\partial\fc^2 < 0$ (Eq.~\eqref{concave}), one has
$\partial\ln\Lhsto/\partial\fc \geqslant 0$ and thus $g(\fc) \geqslant \fc$
for $\fc \in [0, \hat\fc]$; similarly,
$\partial\ln\Lhsto/\partial\fc \leqslant 0$ and $g(\fc) \leqslant \fc$
for $\fc \in [\hat\fc, 1]$.
Because
\[
\frac{\df g}{\df \fc} = \frac{1}{\nI\*\nJ}\*\sum_{i=1}^\nI
\frac{\xi_{i\sep 0} \* \sum_{k=1}^{\nJ} \xi_{i\sep k}}{
  (\sum_{k=0}^{\nJ} \zeta_{i\sep k})^2} > 0,
\]
$g$ is also an increasing function.

Let us consider the case $\fc_0 \in [0, \hat\fc]$.
If $\fc_k \leqslant \hat\fc$, 
$g(\fc_k) \geqslant \fc_k$ and
$g(\fc_k) \leqslant g(\hat\fc) = \hat\fc$. As $g(\fc_k) = \fc_{k+1}$,
$(f_k)_{k\in\mathbb{N}}$ is an increasing sequence
bounded from above by $\hat\fc$: it converges therefore in $[\fc_0, \hat\fc]$.
Because $g$ is continuous and $\hat\fc$ is the only fixed point in this 
interval, $(f_k)_{k\in\mathbb{N}}$ tends to $\hat\fc$.

Similarly, if $\fc_0 \in [\hat\fc, 1]$, $(f_k)_{k\in\mathbb{N}}$ is a decreasing 
sequence converging to $\hat\fc$.

\subsection{One-to-one case}
\label{impl_oto}
All what was said for the several-to-one case still holds 
in the one-to-one case.
Incidentally, as the former is computationally much simpler than the latter,
it is a good idea to compute first $\hat{\vec x}_\sto$ 
and the probabilities 
$\expandafter\hat\Psto(\Assoc_{i\sep j} \mid \COORD \cap \COORD')$:
as $\hat\fc_{\sto}/\nJ > \hat\fc'_{\sto}/\nI$ and 
$\hat\fc_{\oto}/\nJ = \hat\fc'_{\oto}/\nI$, the several-to-one assumption 
is probably correct if $\hat\fc_{\sto}/\nJ \gg \hat\fc'_{\sto}/\nI$;
and if not, one may first test the one-to-several (subscript ``$\ots$''
hereafter)
assumption, i.e. reverse the roles of $\KI$ and $\KJ$ in all the formulae of 
Sect.~\ref{several}, and adopt it if 
$\hat\fc_{\ots}/\nJ \ll \hat\fc'_{\ots}/\nI$. 

Ideally, one would compare the likelihood
of each assumption and adopt the most likely one. While
$\expandafter\hat\Lhsto$ and $\hat\Lh_{\ots}$ are easily computed,
no convenient expression was found for $\expandafter\hat\Lhoto$.
However, if $\ln\expandafter\hat\Lhsto$ and $\ln\hat\Lh_{\ots}$ are of the 
same order, this provides some hint that the one-to-one case (or maybe the 
several-to-several one!) should be considered.
Even then,
$\hat{\vec x}_\sto$ will still 
be a good starting point to find $\hat{\vec x}_\oto$
and there will be no need to compute 
$\expandafter\hat\Poto(\Assoc_{i\sep j} \mid \COORD \cap \COORD')$
for all couples $(i, j)$
such that $\expandafter\hat\Psto(\Assoc_{i\sep j} \mid \COORD \cap \COORD') 
\approx \hat\Prob_{\ots}(\Assoc_{i\sep j} \mid \COORD \cap \COORD') \approx 1$.

The results of Sect.~\ref{fractionoto} are given in terms of 
$\Poto(\Assoc_{i\sep j} \mid \COORD \cap \COORD')$.
The only difficulty is to estimate this probability 
from Eq.~\eqref{Poto}. Because of the combinatorial explosion of the number
of terms, an exact computation is hopeless. 
An approximate value might however be obtained in the following way.

For any $\MI_i$, let $\phi$ be a permutation on $\KI$ 
ordering the elements $\MI_{\phi(1)}$, $\MI_{\phi(2)}$, \textellipsis, 
$\MI_{\phi(\nI)}$
by increasing angular distance to $\MI_i$.
For $j=0$ or $\MJj$ in the neighborhood of $\MI_i$, and for any 
$\ell \in \IE[1, \nI]$,
define
\begin{align}
  \label{Pl}
  \Prob_\ell(\Assoc_{i\sep j} \mid \COORD \cap \COORD') 
  &\equiv \frac{\zeta_{i\sep j}\*
    \sum_{\leftsubstack{j_2=0 \\ j_2\not\in J^{\phi\sep\ast}_1}}^{\nJ}
    \cdots \sum_{\leftsubstack{j_\ell=0 \\ j_\ell\not\in J^{\phi\sep\ast}_{\ell-1}}}^{\nJ}
    \prod_{k=2}^\ell \eta^{\phi\sep\ast}_{k\sep j_k}
  }{
    \sum_{\leftsubstack{j_1=0 \\ j_1\not\in J^\phi_0}}^{\nJ}
    \sum_{\leftsubstack{j_2=0 \\ j_2\not\in J^\phi_1}}^{\nJ}
    \cdots 
    \sum_{\leftsubstack{j_\ell=0 \\ j_\ell\not\in J^\phi_{\ell-1}}}^{\nJ}
    \prod_{k=1}^\ell \eta^\phi_{k\sep j_k}
  },
\end{align}
where 
$J^{\phi\sep\ast}_1 \equiv \{j\} \setminus \{0\}$, 
$J^{\phi\sep\ast}_{k} \equiv (J^{\phi\sep\ast}_{k-1} \cup \{j_k\}) \setminus \{0\}$
for all $k \in \IE[2, \nI]$, $J^\phi_k \equiv J_k$ for all $k$,
\[
\eta^\phi_{k\sep j_k} 
\equiv \frac{\fc\*\xi_{\phi(k)\sep j_k}}{\nJ-\card J^\phi_{k-1}}
\quad \text{and}\quad
\eta^{\phi\sep\ast}_{k\sep j_k} 
\equiv \frac{\fc\*\xi_{\phi(k)\sep j_k}}{\nJ-\card J^{\phi\sep\ast}_{k-1}}
\quad \text{if } j_k > 0, 
\]
and 
$\eta^\phi_{k\sep 0} \equiv \eta^{\phi\sep\ast}_{k\sep 0} 
\equiv \zeta_{\phi(k)\sep 0}$.

As $\phi(1) = i$, $\Prob_1(\Assoc_{i\sep j} \mid \COORD \cap \COORD') = 
\Psto(\Assoc_{i\sep j} \mid \COORD \cap \COORD')$
(cf.~Eq.~\eqref{Psto}): at first order, we obtain the same result as in the
several-to-one case. Since the influence of other $\KI$-sources on the result
decreases very fast with their angular distance to $\MI_i$ and $\MJj$ if 
$\MI_i$ and $\MJj$ are close to each other, 
$\Prob_\ell(\Assoc_{i\sep j} \mid \COORD \cap \COORD')$ should 
rapidly converge to 
$\Prob_{\nI}(\Assoc_{i\sep j} \mid \COORD \cap \COORD') 
= \Poto(\Assoc_{i\sep j} \mid \COORD \cap \COORD')$, 
even for small values of $\ell$.

Because of the recursive sums in Eq.~\eqref{Pl},
the computation must in practice be further
restricted to sources $\MJk$ in the neighborhood of $\MI_i$ and $\MJj$,
as explained in Sect.~\ref{neighbors}.
\appendix
\section{Covariance matrix}
\label{cov}
Let us first remind a few standard results.
The probability that a $q$-dimensional normally distributed random vector
$\vec W$ of mean $\vec\mu$ falls in some domain $\Omega$ is
\[
\Prob(\vec W \in \Omega) = \int_{\vec w \in \Omega} 
\frac{\exp\Bigl(-\frac{1}{2}\*\transpose{[\vec w-\vec\mu]_\BASE}
  \cdot \Gamma_\BASE^{-1} \cdot [\vec w-\vec\mu]_\BASE\Bigr)}{(2\*\pi)^{q/2}
  \*(\det\Gamma_\BASE)^{1/2}}\*\df^q\vec w_\BASE,
\]
where $\BASE \equiv (\vec{\base_1}, \ldots, \vec{\base_q})$ is a basis, 
$\vec w_\BASE \equiv \transpose{(w_1, \dotsc, w_q)}$ 
is the column vector in $\BASE$ of $\vec w = \sum_{i=1}^q w_i\*\vec{\base_i}$, 
$\df^q\vec w_\BASE \equiv \df w_1\times\dotsb\times\df w_q$ and $\Gamma_\BASE$ is the 
covariance matrix of $\vec W$ in $\BASE$.
We note this $\vec W_\BASE \sim\gauss_q(\vec\mu_\BASE, \Gamma_\BASE)$. 

In another basis 
$\BASE' \equiv (\vec{\base'_{\smash[t]{1}}}, \ldots, \vec{\base'_{\smash[t]{q}}})$, 
one has $\vec w_\BASE = \chgbase_{\BASE\rightarrow \BASE'} \cdot \vec w_{\BASE'}$, 
where
$\chgbase_{\BASE\rightarrow \BASE'}$ is the transformation matrix from $\BASE$ to $\BASE'$
(i.e.\ 
$\vec{\base'_{\smash[t]{j}}} = 
\sum_{i=1}^q (\chgbase_{\BASE\rightarrow \BASE'})_{i\sep j}\*\vec{\base_i}$).
Since
$
\df^q\vec w_\BASE = \lvert\det \chgbase_{\BASE\rightarrow \BASE'}\rvert\*\df^q\vec w_{\BASE'}
$
and
\[
\transpose{(\vec w-\vec\mu)_\BASE} \cdot \Gamma_\BASE^{-1} \cdot (\vec w-\vec\mu)_\BASE =
\transpose{(\vec w-\vec\mu)_{\BASE'}} \cdot \Bigl(\chgbase_{\BASE\rightarrow \BASE'}^{-1} \cdot
\Gamma_\BASE \cdot \transpose{[\chgbase_{\BASE\rightarrow \BASE'}^{-1}]}\Bigr)^{-1} \cdot
(\vec w-\vec\mu)_{\BASE'},
\]
one still obtains
\[
\Prob(\vec W \in \Omega) = \int_{\vec w \in \Omega} 
\frac{\exp\Bigl(-\frac{1}{2}\*\transpose{[\vec w-\vec\mu]_{\BASE'}}
  \cdot \Gamma_{\BASE'}^{-1} \cdot [\vec w-\vec\mu]_{\BASE'}\Bigr)}{(2\*\pi)^{q/2}
  \*(\det\Gamma_{\BASE'})^{1/2}}\*\df^q\vec w_{\BASE'},
\]
where
$\Gamma_{\BASE'} = \chgbase_{\BASE\rightarrow \BASE'}^{-1} \cdot \Gamma_\BASE \cdot
\transpose{(\chgbase_{\BASE\rightarrow \BASE'}^{-1})}$ is the covariance matrix 
of $\vec W$ in $\BASE'$.
In the following, $\BASE$ and $\BASE'$ are orthonormal bases, so 
$\chgbase_{\BASE\rightarrow \BASE'}$ is a rotation matrix. From 
$\transpose{\chgbase_{\BASE\rightarrow \BASE'}} = 
\chgbase_{\BASE\rightarrow \BASE'}^{-1}$, one gets
$
\Gamma_{\BASE'} = \transpose{\chgbase_{\BASE\rightarrow \BASE'}} \cdot \Gamma_\BASE \cdot \chgbase_{\BASE\rightarrow \BASE'}
$.

In a common basis, 
for independent random vectors 
$\vec{W_1} \sim\gauss_q(\vec{\mu_1}, \Gamma_1)$ and 
$\vec{W_2} \sim\gauss_q(\vec{\mu_2}, \Gamma_2)$, we have
\[
\vec{W_1} \pm \vec{W_2} \sim\gauss_q(\vec{\mu_1} \pm \vec{\mu_2}, 
\Gamma_1+\Gamma_2).
\]

We now use these results to obtain the covariance matrix of vector
$\vec{r_{i\sep j}} \equiv \vrpj-\vec{r_i}$, where $\vec{r_i}$ and $\vrpj$
are, respectively, the observed positions of source $\MI_i$ of $\KI$ 
and of its counterpart $\MJj$ in $\KJ$. 
We note $\vrzi$ and $\vrpzj$ their true positions.
One has 
\[
\vec{r_{i\sep j}} = (\vrpj - \vrpzj) + 
(\vrpzj - \vrzi) + (\vrzi -\vec{r_i}).
\]
We drop the subscript and the ``prime'' symbol in the following whenever an
expression depends on either $\MI_i$ or $\MJj$ only.

Let $(\vec{u_x}, \vec{u_y}, \vec{u_z})$ be
a direct orthonormal basis, with
$\vec{u_z}$ oriented from the Earth's center $O$ to the North Celestial Pole 
and $\vec{u_x}$ from $O$ to the Vernal Point. 
At a point $\MI$ of right ascension $\alpha$ and declination $\delta$,
a direct orthonormal basis $(\vec{u_r}, \vec{u_\alpha}, \vec{u_\delta})$
is defined by
\begin{align*}
  \vec{u_r} &\equiv \frac{\vec{OM}}{\lVert\vec{OM}\rVert} = 
  \cos\delta\*\cos\alpha\*\vec{u_x} 
  + \cos\delta\*\sin\alpha\*\vec{u_y} + \sin\delta\*\vec{u_z},\\
  \vec{u_\alpha} &\equiv 
  \frac{\partial\vec{u_r}/\partial\alpha}{
    \lVert\partial\vec{u_r}/\partial\alpha\rVert}
  = -\sin\alpha\*\vec{u_x} + \cos\alpha\*\vec{u_y},\\
  \vec{u_\delta} &\equiv
  \frac{\partial\vec{u_r}/\partial\delta}{
    \lVert\partial\vec{u_r}/\partial\delta\rVert}
  = -\sin\delta\*\cos\alpha\*\vec{u_x} - \sin\delta\*\sin\alpha\*\vec{u_y}
  + \cos\delta\*\vec{u_z}.&
\end{align*}

The uncertainty ellipse on the position of $\MI$ is characterized
by the lengths $a$ and $b$ of the semi-major and semi-minor axes,
and by the position angle $\beta$ between the North and the semi-major axis.
Let $\vec{u_a}$ and $\vec{u_b}$
be unit vectors directed respectively along the major and the minor axes, 
and such that 
$(\vec{u_r}, \vec{u_a}, \vec{u_b})$ is a direct orthonormal basis
and 
$\beta \equiv (\widehat{\vec{u_\delta}, \vec{u_a}})$ is in
$[0, \pi]$ when counted eastward.
In the plane oriented by $+\vec{u_r}$, 
\[
\chgbase_{(\vec{u_a}{,}\, \vec{u_b})\rightarrow(\vec{u_\alpha}{,}\, \vec{u_\delta})} = 
\Left(\begin{matrix}
  \sin\beta & \cos\beta \\
  -\cos\beta & \sin\beta
\end{matrix}\Right) \equiv \Rot(\beta),
\]
since $(\vec{u_\alpha}, \vec{u_\delta})$ is obtained from $(\vec{u_a}, \vec{u_b})$
by a $(\beta-\pi/2)$-counterclockwise rotation.
As\footnote{We seize this opportunity to correct equations (A.8) to (A.11)
  of \citet{Pineau}: $a$ and $b$ should be replaced by their squares in these 
  formulae.} 
\[
\Gamma_{(\vec{u_a}{,}\, \vec{u_b})} = 
\Left(\begin{matrix} 
  a^2 & 0 \\
  0 & b^2
\end{matrix}\Right) \equiv \Diag\bigl(a^2, b^2\bigr),%
\]
one has
$
\Gamma_{(\vec{u_\alpha}{,}\, \vec{u_\delta})} = 
\transpose{\Rot}(\beta) \cdot \Diag\bigl(a^2, b^2\bigr) \cdot \Rot(\beta)
$.

As noticed by \citet{Pineau}, for sources close to the Poles,
$(\vec{u_{\alpha_i}}, \vec{u_{\delta_i}}) \not\approx 
(\uapj, \udpj)$, so one needs to define a common basis.
We use the same basis as them, noted $(\vec t, \vec n)$ below. While
the results we get are intrinsically the same, some people may find
our expressions more convenient.

Denote $\psi \equiv (\widehat{\vec{u_{r_i}}, \urpj}) \in [0, \pi]$ 
the angular distance between $\MI_i$ and $\MJj$,
and
$
\vec n \equiv
\vec{u_{r_i}} \times \urpj/\lVert\vec{u_{r_i}} \times \urpj\rVert
$ a unit vector perpendicular to the plane $(O, \MI_i, \MJj)$.
One has 
$\vec{u_{r_i}}\cdot\urpj = \cos\psi$,
so 
\begin{align}
  \label{psi}
  \psi = \arccos(\cos\delta_i\*\cos\deltapj\*\cos[\alphapj-\alpha_i] 
  + \sin\delta_i\*\sin\deltapj),
\end{align}
and
$\lVert\vec{u_{r_i}} \times \urpj\rVert = \sin\psi$.

Let $\gamma_i \equiv (\widehat{\vec n, \vec{u_{\delta_i}}})$ 
and $\gammapj \equiv (\widehat{\vec n, \udpj})$ 
be angles oriented clockwise around
$+\vec{u_{r_i}}$ and $+\urpj$, respectively.
Angle $\gamma_i$ is fully determined by following expressions:
\begin{align*}
  \cos\gamma_i &= \vec n\cdot \vec{u_{\delta_i}} = 
  \frac{1}{\sin\psi}\*(\vec{u_{r_i}} \times \urpj) \cdot \vec{u_{\delta_i}} =
  \frac{1}{\sin\psi}\*(\vec{u_{\delta_i}} \times \vec{u_{r_i}}) \cdot \urpj =
  \frac{1}{\sin\psi}\*\vec{u_{\alpha_i}}\cdot \urpj\\
  &= \frac{\cos\deltapj\*\sin(\alphapj-\alpha_i)}{\sin\psi}; \\
  \sin\gamma_i &= -\vec n\cdot \vec{u_{\alpha_i}} = 
  -\frac{1}{\sin\psi}\*(\vec{u_{r_i}} \times \urpj) \cdot \vec{u_{\alpha_i}} =
  -\frac{1}{\sin\psi}\*(\vec{u_{\alpha_i}} \times \vec{u_{r_i}}) \cdot \urpj =
  \frac{1}{\sin\psi}\*\vec{u_{\delta_i}}\cdot \urpj\\
  &= 
  \frac{\cos\delta_i\*\sin\deltapj - 
    \sin\delta_i\*\cos\deltapj\cos(\alphapj-\alpha_i)}{\sin\psi}.
\end{align*}
Similarly, 
\[
\cos\gammapj = \frac{\cos\delta_i\*\sin(\alphapj-\alpha_i)}{\sin\psi}
\quad\text{and}\quad
\sin\gammapj = \frac{\cos\delta_i\*\sin\deltapj\cos(\alphapj-\alpha_i) 
  - \sin\delta_i\*\cos\deltapj}{\sin\psi}.
\]
Note that determining $\gamma_i$ and $\gammapj$ themselves might slow down the 
computations: for instance, only the
sines and cosines of $\beta_i$ and $\gamma_i$ are of interest in the matrices 
$\Rot(\beta_i + \gamma_i)$ used hereafter, as is obvious from the expansion of 
$\sin(\beta_i + \gamma_i)$ and $\cos(\beta_i + \gamma_i)$. 
The same holds for $\Rot(\betapj + \gammapj)$ and other matrices.

Let $\vec t \equiv \vec n \times \vec{u_{r_i}}$: $\vec t$ is a unit
vector tangent
in $\MI_i$ to the minor arc of great circle going from $\MI_i$ to $\MJj$.
Project the sphere on the plane $(\MI_i, \vec t, \vec n)$ tangent 
to the sphere in $\MI_i$ (which specific projection does not matter since we
consider only $\KJ$-sources in the neighborhood of $\MI_i$): 
one has $\vec{r_{i\sep j}} \approx \psi\*\vec t$,
and the basis $(\vec t, \vec n)$ is obtained from $(\vec{u_a}, \vec{u_b})$
by a $(\beta+\gamma-\pi/2)$-counterclockwise rotation around $+\vec{u_r}$, so,
in $(\vec t, \vec n)$,
\[
\Gamma_i = 
\transpose{\Rot}(\beta_i+\gamma_i) \cdot \Diag\bigl(a_i^2, b_i^2\bigr) 
\cdot \Rot(\beta_i+\gamma_i)
\quad \text{and} \quad
\Gammapj = 
\transpose{\Rot}(\betapj+\gammapj) \cdot 
\Diag\bigl(a_{\smash[t]{j}}'^2, b_{\smash[t]{j}}'^2\bigr) 
\cdot \Rot(\betapj+\gammapj).
\]
As $\vec{r_i} \sim \gauss_2(\vec 0,\Gamma_i)$ and 
$\vrpj \sim \gauss_2(\vec 0,\Gammapj)$, 
one has
$\vec{r_{i\sep j}} \sim \gauss_2(\vec 0, \Gamma_{i\sep j})$
if the true positions are identical, 
where
$\Gamma_{i\sep j} \equiv \Gamma_i + \Gammapj$. 

If the positional
uncertainty on $\MI_i$ is unknown, one may assume that
$\Gamma_i = \sigma^2\*\Diag(1, 1)$, with the same $\sigma$
for all $\KI$-sources, and derive $\sigmatot \equiv \sigma$ 
by maximizing the likelihood to
observe the distribution of $\KI$-sources given that of $\KJ$-sources
(see Sects.~\ref{fractionsto} and~\ref{fractionoto}). For a galaxy, however,
the positional uncertainty on its center is likely to increase
with its size. If the position angle $\theta_i$ (counted eastward from
the North) and the major and minor diameters $D_i$ and $d_i$ 
of the best-fitting ellipse of some isophote are known 
for $\MI_i$
(for instance, parameters $\textsc{\large pa}$, $\Dvc$ and $\dvc \equiv \Dvc/R_{25}$
taken from the \textsc{\large rc}3 catalog \citep{RC3} or HyperLeda \citep{HYPERLEDA}),
one may model $\Gamma_i$ as
\[
\Gamma_i = \transpose{\Rot}(\gamma_i+\theta_i) \cdot
\Diag\bigl(\sigma^2 + [\nu\*D_i]^2, \sigma^2 + [\nu\*d_i]^2\bigr)
\cdot \Rot(\gamma_i+\theta_i)
= \sigma^2\*\Diag(1, 1) + \nu^2\*\transpose{\Rot}(\gamma_i+\theta_i) \cdot
\Diag\bigl(D_i^2, d_i^2\bigr) \cdot \Rot(\gamma_i+\theta_i),
\]
and derive both $\sigmatot \equiv \sigma$ and $\nutot \equiv \nu$ 
from the maximum likelihood.
Such a technique might indeed be used to estimate the accuracy of coordinates
in some catalog (see \citet{PP} for another method).

If the positional uncertainty on $\MJj$ is also unknown,
one can put 
\[
\Gammapj = 
\sigma'^2\*\Diag(1, 1) + \nu'^2\*\transpose{\Rot}(\gammapj+\thetapj) \cdot
\Diag\bigl(D^2, d^2\bigr) \cdot \Rot(\gammapj+\thetapj)
\]
with the same $\sigma'$ and $\nu'$ for all $\KJ$-sources. 
As $\gammapj + \thetapj = \gamma_i + \theta_i$, 
only $\sigmatot \equiv \bigl(\sigma^2+\sigma'^2\bigr)^{1/2}$ and
$\nutot \equiv \bigl(\nu^2+\nu'^2\bigr)^{1/2}$ may be obtained%
\footnote{%
  However, as noticed by \citet{dVH} in a different context, 
  if three samples with unknown
  uncertainties $\sigma_i$ ($i \in \IE[1, 3]$) are available
  and if the $\sigma_{i\sep j} \equiv (\sigma_i^2+\sigma^2_j)^{1/2}$
  may be estimated for all the pairs $(i, j)_{j\neq i} \in \IE[1, 3]^2$, 
  as in our case, then $\sigma_i$ may be determined for 
  each sample.
  \citet{PP} used this technique to compute the accuracy of galaxy 
  coordinates.
}
\ from the maximum likelihood, not $\sigma$, $\sigma'$, $\nu$ or $\nu'$.

A similar technique can be applied if the true centers of a source in
$\KI$ and of its counterpart in $\KJ$ may differ.
This might be in particular useful when associating galaxies from an
optical catalog and from a ultraviolet or far-infrared catalog, 
because, while
the optical is dominated by smoothly-distributed evolved stellar populations,
the ultraviolet and the far-infrared mainly trace star-forming regions.
Observations of galaxies by \citet{Kuchinski}
have indeed shown that galaxies are very patchy in the ultraviolet,
and the same has been observed in the far-infrared.
As the angular distance between the true centers should increase with the 
size of the galaxy, one may model this as
$\vrpzj-\vrzi\sim\gauss_2(\vec 0, \Gamma_0),$
where $\Gamma_0 = \nu_0^2\*
\transpose{\Rot}(\gamma_i+\theta_i) \cdot \Diag\bigl(D_1^2, d_1^2\bigr) \cdot
\Rot(\gamma_i+\theta_i)$.

In the most general case,
\[
\vec{r_{i\sep j}} 
\sim\gauss_2(\vec 0, \Gamma_{i\sep j}),
\]
with $\Gamma_{i\sep j} \equiv \Gamma_i+\Gammapj+\Gamma_0$.
Once again, if $\sigma$, $\sigma'$, $\nu$, $\nu'$ and $\nu_0$
are unknown, the quantities 
$\sigmatot \equiv \bigl(\sigma^2+\sigma'^2\bigr)^{1/2}$ and 
$\nutot \equiv \bigl(\nu^2+\nu'^2+\nu_0^2\bigr)^{1/2}$ 
may be determined as indicated in Sects.~\ref{fractionsto} 
and~\ref{fractionoto}.
\begin{acknowledgements}
  The initial phase of this work took place
  at the NASA/Goddard Space Flight Center, under the 
  supervision of Eli Dwek, and was supported by the National Research 
  Council through the Resident Research Associateship Program.
  We acknowledge them sincerely.
\end{acknowledgements}
\bibliographystyle{aa}
\bibliography{references}

\begin{thebibliography}{17}
\expandafter\ifx\csname natexlab\endcsname\relax\def\natexlab#1{#1}\fi

\bibitem[{{Budav{\'a}ri} \& {Szalay}(2008)}]{BS}
{Budav{\'a}ri}, T. \& {Szalay}, A.~S. 2008, \apj, 679, 301

\bibitem[{{Condon} {et~al.}(1975){Condon}, {Balonek}, \& {Jauncey}}]{Condon}
{Condon}, J.~J., {Balonek}, T.~J., \& {Jauncey}, D.~L. 1975, \aj, 80, 887

\bibitem[{{de Ruiter} {et~al.}(1977){de Ruiter}, {Arp}, \& {Willis}}]{DeRuiter}
{de Ruiter}, H.~R., {Arp}, H.~C., \& {Willis}, A.~G. 1977, \aaps, 28, 211

\bibitem[{{de Vaucouleurs} {et~al.}(1991){de Vaucouleurs}, {de Vaucouleurs},
  {Corwin}, {Buta}, {Paturel}, \& {Fouqu{\'e}}}]{RC3}
{de Vaucouleurs}, G., {de Vaucouleurs}, A., {Corwin}, Jr., H.~G., {et~al.}
  1991, {Third Reference Catalogue of Bright Galaxies}, ed. {de Vaucouleurs,
  G., de Vaucouleurs, A., Corwin, H.~G., Jr., Buta, R.~J., Paturel, G., \&
  Fouqu{\'e}, P.}

\bibitem[{{de Vaucouleurs} \& {Head}(1978)}]{dVH}
{de Vaucouleurs}, G. \& {Head}, C. 1978, \apjs, 36, 439

\bibitem[{{Kendall} \& {Stuart}(1979)}]{KS}
{Kendall}, M. \& {Stuart}, A. 1979, {The advanced theory of statistics. Vol.2:
  Inference and relationship}, ed. {Kendall, M.~\& Stuart, A.}

\bibitem[{{Kuchinski} {et~al.}(2000){Kuchinski}, {Freedman}, {Madore},
  {Trewhella}, {Bohlin}, {Cornett}, {Fanelli}, {Marcum}, {Neff}, {O'Connell},
  {Roberts}, {Smith}, {Stecher}, \& {Waller}}]{Kuchinski}
{Kuchinski}, L.~E., {Freedman}, W.~L., {Madore}, B.~F., {et~al.} 2000, \apjs,
  131, 441

\bibitem[{{Moshir} {et~al.}(1993){Moshir}, {Copan}, {Conrow}, {McCallon},
  {Hacking}, {Gregorich}, {Rohrbach}, {Melnyk}, {Rice}, \& {Fullmer}}]{FSC}
{Moshir}, M., {Copan}, G., {Conrow}, T., {et~al.} 1993, VizieR Online Data
  Catalog, 2156, 0

\bibitem[{{Moshir} {et~al.}(1992){Moshir}, {Kopman}, \& {Conrow}}]{FSS}
{Moshir}, M., {Kopman}, G., \& {Conrow}, T.~A.~O. 1992, {IRAS Faint Source
  Survey, Explanatory supplement version 2}, ed. {Moshir, M., Kopman, G., \&
  Conrow, T.~A.~O.}

\bibitem[{{Paturel} {et~al.}(1995){Paturel}, {Bottinelli}, \&
  {Gouguenheim}}]{LEDA}
{Paturel}, G., {Bottinelli}, L., \& {Gouguenheim}, L. 1995, Astrophysical
  Letters and Communications, 31, 13

\bibitem[{{Paturel} \& {Petit}(1999)}]{PP}
{Paturel}, G. \& {Petit}, C. 1999, \aap, 352, 431

\bibitem[{{Paturel} {et~al.}(2003){Paturel}, {Petit}, {Prugniel}, {Theureau},
  {Rousseau}, {Brouty}, {Dubois}, \& {Cambresy}}]{HYPERLEDA}
{Paturel}, G., {Petit}, C., {Prugniel}, P., {et~al.} 2003, VizieR Online Data
  Catalog, 7237, 0

\bibitem[{{Pineau} {et~al.}(2011){Pineau}, {Motch}, {Carrera}, {Della Ceca},
  {Derri{\`e}re}, {Michel}, {Schwope}, \& {Watson}}]{Pineau}
{Pineau}, F.-X., {Motch}, C., {Carrera}, F., {et~al.} 2011, \aap, 527, A126

\bibitem[{{Prestage} \& {Peacock}(1983)}]{Prestage}
{Prestage}, R.~M. \& {Peacock}, J.~A. 1983, \mnras, 204, 355

\bibitem[{{Rutledge} {et~al.}(2000){Rutledge}, {Brunner}, {Prince}, \&
  {Lonsdale}}]{Rutledge}
{Rutledge}, R.~E., {Brunner}, R.~J., {Prince}, T.~A., \& {Lonsdale}, C. 2000,
  \apjs, 131, 335

\bibitem[{{Sutherland} \& {Saunders}(1992)}]{SS}
{Sutherland}, W. \& {Saunders}, W. 1992, \mnras, 259, 413

\bibitem[{{Wolfram}(1996)}]{Mathematica}
{Wolfram}, S. 1996, {The Mathematica book}, ed. {Wolfram, S.}

\end{thebibliography}
\end{document}